\documentclass[journal]{IEEEtran}
\usepackage{times}
\usepackage{makecell}
\usepackage{epsfig}
\usepackage{graphicx}
\usepackage{amsmath}
\usepackage{amssymb}
\usepackage{wrapfig}
\usepackage[ruled, linesnumbered]{algorithm2e}
\usepackage{xcolor}
\usepackage{bbm}
\usepackage{algorithmic}
\usepackage{subcaption}
\usepackage{microtype}      % microtypography
\usepackage{makecell}
\usepackage{multirow}
\usepackage{mathtools}
\usepackage{algorithmic}
\usepackage{booktabs} 
\usepackage{diagbox}
\usepackage{setspace}
\usepackage{threeparttable}
\usepackage[pagebackref=true,breaklinks=true,colorlinks,bookmarks=false]{hyperref}
\ifCLASSINFOpdf
\else
\fi
\begin{document}
\title{Learning from Synthetic Data for Opinion-free Blind Image Quality Assessment in the Wild}

\author{Zhihua~Wang,
        Zhi-Ri Tang,
        Jianguo Zhang,~\IEEEmembership{Senior Member,~IEEE},
        and Yuming Fang,~\IEEEmembership{Senior Member,~IEEE}% <-this % stops a space
\thanks{Zhihua Wang is with the Department
of Computer Science, City University of Hong Kong,
Hong Kong, also with the Department of Computer Science and Engineering, Southern University of Science and Technology, Shenzhen, China (e-mail: zhihua.wang@my.cityu.edu.hk).}% <-this % stops a space
\thanks{Zhi-Ri Tang is with the School of Physics and Technology, Wuhan University, Wuhan, China (e-mail: GerinTang@163.com).}
\thanks{Jianguo Zhang is with the Department of Computer Science and Engineering, Southern University of Science and Technology, Shenzhen, China (e-mail: zhangjg@sustech.edu.cn).}% <-this % stops a space
\thanks{Yuming~Fang is with the School of Information Management, Jiangxi University of Finance and Economics, Nanchang China (e-mail: fa0001ng@e.ntu.edu.sg).}% <-this % stops a space
}
%\thanks{Manuscript received April 19, 2015; revised August 26, 2015.}}


\markboth{}%
{Shell \MakeLowercase{\textit{et al.}}: Bare Demo of IEEEtran.cls for IEEE Journals}



\maketitle

\begin{abstract}
%Recently, increasing interest has been drawn in leveraging deep convolutional neural networks (CNNs) for blind image quality assessment (BIQA).
%\textcolor{red}{Early attempts of $opinion$-$free$ convolutional neural network (CNN) based blind image quality assessment (BIQA) models are mainly optimized and tested on synthetically-degraded images, generalizing poorly to authentically-distorted images captured in the wild due to their distributional shift. To alleviate this challenge, in this paper, we first present a unified method of training BIQA models targeting the quality assessment of $authentically$-$distorted$ images without relying on human opinion scores.}
Nowadays, most existing blind image quality assessment (BIQA) models 1) are developed for synthetically-distorted images and often generalize poorly to authentic ones; 2) heavily rely on human ratings, which are prohibitively labor-expensive to collect. Here, we propose an $opinion$-$free$ BIQA method that learns from synthetically-distorted images and multiple agents to assess the perceptual quality of authentically-distorted ones captured in the wild without relying on human labels.
Specifically, we first assemble a large number of image pairs from synthetically-distorted images and use a set of full-reference image quality assessment (FR-IQA) models to assign pseudo-binary labels of each pair indicating which image has higher quality as the supervisory signal. We then train a convolutional neural network (CNN)-based BIQA model to rank the perceptual quality, optimized for consistency with the binary labels. Since there exists domain shift between the synthetically- and authentically-distorted images, an unsupervised domain adaptation (UDA) module is introduced to alleviate this issue. Extensive experiments demonstrate the effectiveness of our proposed $opinion$-$free$ BIQA model, yielding state-of-the-art performance in terms of correlation with human opinion scores, as well as gMAD competition. Codes will be made publicly available upon acceptance.%: https://github.com/wangzhihua520/OF\_BIQA.git.
\end{abstract}


% \fi
\IEEEpeerreviewmaketitle

\section{Introduction}
\IEEEPARstart 
{D}{igital} images have become ubiquitous in almost every aspect of our life. In those tasks from image acquisition, compression, transmission, to storage, etc., there may inevitably exist degradations of image quality, leading to unsatisfactory visual experience \cite{wang2004image}. Therefore, it is of great significance to build accurate image quality assessment (IQA) methods to maintain, control, and boost the perceptual quality of images. Based on the accessibility of original reference images, existing IQA methods can be categorized into \textit{full}-reference IQA (FR-IQA) \cite{wang2004image, wang2003multiscale, zhang2014vsi}, \textit{reduced}-reference IQA (RR-IQA) \cite{wang2011} and \textit{no}-reference/blind IQA (NR-IQA/BIQA) \cite{ma2017dipiq, ma2017end, ma2019blind, zhang2019learning}. In the existing literature, it has been demonstrated that FR-IQA methods usually show better generalizability and robustness compared to \textit{blind} IQA (BIQA) models \cite{pan2018blind}. However, in many practical settings, reference images are not often available (or may not even exist), thus limiting the use of reference-based IQA methods. Blind IQA (BIQA) models are designed to assess the image quality without the need of reference images.  
%Such a model should be appealing in practical applications because humans are able to estimate the perceptual quality of distorted images without comparison to any pristine reference. Presumably human achieve this ability through long-time evolutionary and developmental processes, learned to preferentially represent and recognize "natural" visual images. Due to the lack of reference information, BIQA is more challenging than other types of approaches. 
\begin{figure}[t]
	\centering
	% \captionsetup{justification=centering}
	\subfloat[Synthetic overexposure]{\includegraphics[width=0.24\textwidth]{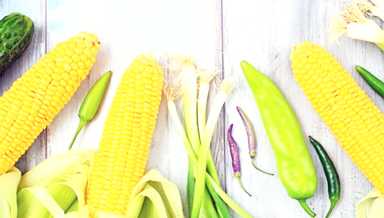}}\hskip.25em
	\subfloat[Synthetic motion blur]{\includegraphics[width=0.24\textwidth]{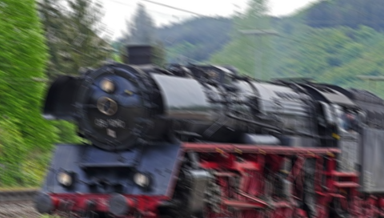}}\\
	\subfloat[Authentic overexposure]{\includegraphics[width=0.24\textwidth]{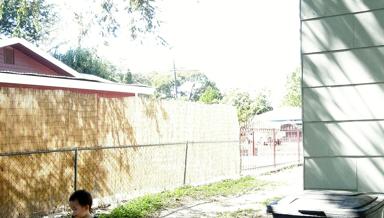}}\hskip.25em
	\subfloat[Authentic motion blur]{\includegraphics[width=0.24\textwidth]{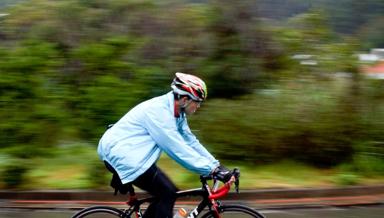}}\\
	\caption{Representative samples of synthetic distortions from KADID-10k \cite{lin2019kadid} and authentic distortions from KonIQ-10k \cite{hosu2020koniq}. \textbf{(a)}-\textbf{(b)} synthetically-distorted images, showing uniformity of distortions . \textbf{(c)}-\textbf{(d)} authentically-distorted images, exhibiting non-uniformity.
	%Images in the right panel has more diversity then those on the left panel.
% 	In this setting, the selected gMAD images have limited diversity, most of which are underexposure or striped images. 
% 	The predicted values are mapped onto the CLIVE MOS scale, with a higher number indicating better perceptual quality. 
	}
	\label{fig:diff_syn_aut}
\end{figure}

In principle, BIQA methods first extract features from an image and then map them to a score, indicating its perceptual quality. The key to high performances of BIQA methods lies in the feature representations. Conventional BIQA methods are designed to make use of handcrafted features. Spatially normalized coefficients~\cite{mittal2012no} and codebook-based representations~\cite{ye2012unsupervised} are two examples that have demonstrated impressive performance on common distortion types~\cite{sheikh2006statistical,ma2016waterloo}. Recently, deep neural networks (DNNs) have shown their exceptional capability in the task of automatic learning feature representations, % Instead of carefully designing handcrafted representations, DNNs can learn how to extract feature representations from the inputs.
 and they have been applied in many BIQA models \cite{ma2017end, gu2014deep, zhang2018blind, ma2017dipiq, pan2018blind}, achieving much better performances on existing IQA benchmarks. However, the successful training of such a model with millions of parameters would require massive quality annotations in the form of mean human-opinion scores (MOSs), which are largely lacking until now \cite{lin2020deepfl} due to the fact that annotating a large-scale IQA image dataset is prohibitively labor-expensive and costly \cite{liu2017rankiqa} (see TABLE \ref{tab:dataset} for a list of datasets). One way to handle this issue is to learn from synthetic data with reliable pseudo labels that can be collected automatically at low cost, instead of training BIQA models on scored images \cite{liu2017rankiqa, ye2014beyond, ma2019blind}. 
 
%targeting the objective perceptual quality assessment in the wild 
\begin{table*}
\caption{Summary of the characteristics of different public authentically-distorted IQA datasets}
\centering
\begin{threeparttable}
\begin{tabular}{l|cccccc}
\toprule
dataset & Year & \# of distorted images & \# of annotations & \# of subjects & Score range  & Methodology\tnote{1} \\
\hline
BID~\cite{ciancio2010no} & $2011$ & $586$ & $6,446$ & $180$ & $[0,5]$ & SS-CQR\\
CID2013~\cite{virtanen2014cid2013} & $2013$ & $480$ & $14,260$ & $188$ & $[0,100]$ & ACR-DR\\
LIVE Challenge~\cite{ghadiyaram2015massive}  & $2015$ & $1,162$  & $>0.35$ million & $>8,100$ & $[0, 100]$ & SS-CQR-CS\\
KonIQ-10k~\cite{hosu2020koniq} & $2018$ & $10,073$ & $1.2$ million & 1,459 & $[1, 5]$ & SS-ACR-CS\\
SPAQ~\cite{fang2020cvpr} & $2020$ & $11,125$  & $>0.16$ million & $>600$ & $[0, 100]$ &  SS-CQR\\
\bottomrule
\end{tabular}
\begin{tablenotes}
    \item[1] SS: Single stimulus. CQR: Continuous quality rating. ACR: Absolute category rating. CS: Crowdsourcing. DR: Dynamic reference.
\end{tablenotes}
\end{threeparttable}
\label{tab:dataset}
\end{table*}

Many studies \cite{mu2020learning, varol2017learning} show that by learning from synthetic images, models could achieve supreme performance. Compared to collecting real-world datasets requiring a large number of human annotations, datasets of synthetic images offer several unique advantages. First, collecting a large scale of synthetic images with rich (pseudo) supervisory signals in a lab is much easier and cheaper than annotating a dataset of the same scale captured in the wild. Second, real-world datasets may suffer from the problems of data imbalance and long-tail in distributions; those could be avoided by balancing different categories when synthesizing datasets. In IQA setting, Liu. $et~al$ \cite{liu2017rankiqa} explored to generate a large-scale ranked image dataset, where two images in a pair are ranked relatively w.r.t different distortion levels but of the same distortion type. Therefore the \textit{cross-distortion-type} ranking information is absent in this study. Ye. $et~al$ \cite{ye2014beyond} and Ma. $et~al$ \cite{ma2019blind} proposed to train a model using a dataset by including synthetically-distorted images and corresponding pseudo labels predicted by a group of FR-IQA models, achieving excellent performance on synthetic data. Although 
such methods can also be used for the quality assessment of images captured in the \textit{wild}, they suffer from significant performance decreasing on authentic distortions due to the inevitable \textit{shift} in data distribution between synthetic and real data. 
%To alleviate the performance dropping, we should  need to a \textit{unified} method of training $opinion$-$free$ (a setting where no subjective opinions are used for training) BIQA model by effectively learning from synthetically-distorted images without using any human opinion scores generalizing exceedingly well to IQA in the \textit{wild}.   % pursue this strategy to learn BIQA models from the synthetically-distorted images targeting to qualify the quality of the authentically-degraded images. 

In general, assessing the quality of images degraded by pre-defined \textit{synthetic} distortions, $ e.g.,$ Gaussian blur, JPEG compression artifacts, JPEG 2000 compression artifacts, Gaussian noise, is much easier than assessing real-world ones. Images captured in the wild are usually distorted by a complex mixtures of \textit{multiple} distortions \cite{ghadiyaram2015massive}, which are hard to be well-simulated by man-made distortions \cite{fang2020cvpr, ghadiyaram2015massive, hosu2020koniq}. Fig.\ref{fig:diff_syn_aut} illustrates some representative examples of synthetically- and authentically-distorted images. It could easily observe that synthetic distortions are usually manipulated globally, while authentic ones may be non-uniform~\cite{sun2021blind}. This is a typical example exhibiting the existence of distributional shifts between the simulated and the real-world distortions. Therefore, models trained on synthetic distortions usually perform poorly when generalizing to assess realistic ones. 
To compensate for the degradation in performance, we need to let knowledge learning from synthetically-distorted domains adapt to authentically-distorted ones. Domain adaptation \cite{ganin2016domain, sun2016deep, tzeng2017adversarial, tzeng2014deep} is a plausible way to alleviate this challenge, which often provides an attractive option where labeled training data are enough but lacking annotations for targeting (test) data. 
Until now, rare studies discuss it in the setting of IQA.

% Domain adaptation \cite{tzeng2017adversarial, tzeng2014deep} is a promising way to address this issue, which often provides an attractive option where labeled training data are enough but lacking annotations for targeting (test) data \cite{liebelt2010multi, vazquez2013virtual, sun2014virtual}. \textbf{(we might need to compare with different domain adaptation methods.)}

%a \textit{unified} method of training $opinion$-$free$ (a setting where no subjective opinions are used for training) BIQA model by effectively learning from synthetically-distorted images without using any human opinion scores generalizing exceedingly well to IQA in the \textit{wild}.   % pursue this strategy to learn BIQA models from the synthetically-distorted images targeting to qualify the quality of the authentically-degraded images.

In this paper, we explore how to learn a deep convolutional neural network (CNN) based $opinion$-$free$ BIQA model for the quality assessment of authentically-distorted images without reliance on any human scores. Technically, we first generate a large number of image pairs, and for each pair, multiple FR-IQA methods (for clarity, we terms them as \textit{agents} in context.) are used to compute pseudo-binary labels indicating which of the two images is of higher quality. We then train a CNN-based model to compute a quality score by fitting these pseudo-binary labels, using a pairwise learning-to-rank (L2R) strategy \cite{burges2005learning}. It was noted that there are large distributional shifts between synthetically-distorted images and real-world images degraded by authentic distortions, which seriously hamper the generalization performance of models trained on synthetic data when handling authentic images. To address this challenge, we further introduce unsupervised domain adaptation (UDA) techniques \cite{ganin2016domain, sun2016deep, tzeng2017adversarial, tzeng2014deep} to close the domain gap between synthetic images (source domain) and authentic images (target domain). Specifically, for feature extraction, we design a unit backboned by domain adversarial neural networks to learn discriminative and domain-invariant features by exploiting adversarial learning between a feature extractor and a domain discriminator. % Cross-entropy is a classical and standard loss function for domain classifier \cite{de2005tutorial}, where even easily classifiable examples result in a non-negligible loss \cite{saito2018strong}. 
In our design, we want the domain discriminator to focus on hard-to-classify examples while assigning less importance to the easy-to-classify ones during training. Here easy-to-classify and hard-to-classify refer to images being easy and hard to distinguish by domain discriminator from synthetically-distorted or authentically-distorted. 
An adaptive loss is introduced to maneuver this by adding a modulating factor to the cross-entropy loss to put a large weight on hard samples. Moreover, we further explore the usage of domain mixup, which can facilitate a more continuous and linear feature distribution in the latent space with low domain shift, to enhance the performance of adaptive BIQA model % \cite{xu2020adversarial,yan2020improve}
.

In summary, our contributions are three-fold. First, a unified learning framework is firstly proposed to learn computational $opinion$-$free$ BIQA models from synthetically-distorted images for the BIQA in the wild without using any human scores, where the supervised signals are rated by a group of FR-IQA models. Second, a simple and easy-to-implement yet effective UDA method, incorporating an adaptive weight loss and domain mixup, to reduce the distributional shift between synthetically-distorted images and the authentically-distorted ones captured in the wild. To the best of our knowledge, our work is the first to exploit adversarial UDA for IQA. Third, extensive experiments on two large-scale realistic IQA datasets demonstrate our proposed method achieves state-of-the-art performance when being evaluated using both human opinion scores as well as gMAD competition.

The rest of this paper is organized as follows. Section II reviews previous works that are closely related to ours. Section III details the algorithm design. Section IV presents the extensive experiment results, and Section V concludes the paper.

\section{Related Works}
In this section, we review previous works that are closely related to ours, including existing approaches to IQA, BIQA in the wild and existing $opinion$-$free$ BIQA models.

\subsection{Existing Approaches to IQA}
long-standing research topic for over 50 years \cite{mannos1974effects}. High quality evaluation mainly pertains to FR-IQA methods, requiring the distorted image along with its pristine counterpart for assessment, $e.g.,$ SSIM~\cite{wang2004image}, VSI~\cite{zhang2014vsi},  MAD~\cite{larson2010most}. However, 
this kind of method is different from the quality-perceiving of humans, which never requires the information of reference image~\cite{wandell1995foundations}. Strictly speaking, FR-IQA metrics are the way of evaluating the \textit{fidelity} of distorted images~\cite{li2002blind}. BIQA models \cite{moorthy2011blind, mittal2013making, ma2017end} do not rely on the access to reference images for quality evaluation. 
Early attempts of BIQA reckon on hand-crafted features, $e.g.,$ the natural scene statistics (NSS) \cite{mittal2011blind, moorthy2011blind, mittal2012no} and Bag-of-Words (BoW) \cite{ye2012unsupervised,xu2016blind}. In recent years, data-driven methods, such as CNN-based BIQAs~\cite{bosse2017deep, ma2017end}, have outperformed conventional BIQA models based on hand-crafted features in terms of correlation numbers~\cite{wang2004image} and gMAD competition~\cite{ma2018group}. However, these methods mainly rely upon large-scale human-rated datasets.

\subsection{BIQA in the Wild}

In recent years, driven by the popularity and advance of portable cameras, mobile phones, as well as photo-centric social Apps, massive authentically-distorted images appear in our lives \cite{fang2020cvpr, ghadiyaram2015massive, mittal2013making}. Most of these images are captured in the wild by casual, inexpert consumers and are easily distorted by lighting, exposure, noise sensitivity \cite{ghadiyaram2015massive}. Normally, authentically-distorted datasets consist of much more diverse contents and complex distortions compared to synthetically-distorted datasets. Frequently used benchmarks guiding the study of authentic distortions include BID \cite{ciancio2010no}, CID2013 \cite{virtanen2014cid2013}, LIVE Challenge \cite{ghadiyaram2015massive}, KonIQ-10k \cite{hosu2020koniq} and SPAQ \cite{fang2020cvpr}, where KonIQ-10k and SPAQ are two relative larger datasets each containing more than $10,000$ human-scored images.

The majority of existing BIQA models can be used for the evaluation of both synthetic and authentic distortions.
Due to the distribution shift between them~\cite{zhang2019learning}~(see Fig. \ref{fig:diff_syn_aut} as an example), models trained on synthetically-distorted images suffer from performance decreasing when generalizing to authentically-distorted ones. For examples, NIQE \cite{mittal2013making} obtained a performance with correlation numbers reaching to $0.91$ on LIVE \cite{sheikh2006image} but only $0.48$ on LIVE Challenge \cite{ghadiyaram2015massive}. % Independent learning and testing on synthetic or authentic distortions is a plausible way to maintain performance.
Based on the assumption of Thurstone's model, Zhang $et~al.$ \cite{zhang2020uncertainty} proposed a unified L2R framework to train the BIQA model on a combination of synthetic and authentic datasets simultaneously, showing excellent performance on both distortions. Domain adaptation \cite{sun2016deep, tzeng2017adversarial, tzeng2014deep} is another potential way to alleviate the influence of distribution shift between synthetic and authentic distortions. In the IQA setting, Chen $et~al.$ \cite{chen2020no} developed the first Maximum Mean Discrepancy (MMD)-based domain adaptation for the quality assessment of screen content images (SCIs). However, different from ours, their methods were supervised and designed for human-rated data; the problem setting of ours is much more challenging. This is by far the first and only study, to the best of our knowledge, that involves domain adaptation in IQA setting. 

\begin{figure*}[t]
	\begin{center}
		\includegraphics[width=0.9\linewidth]{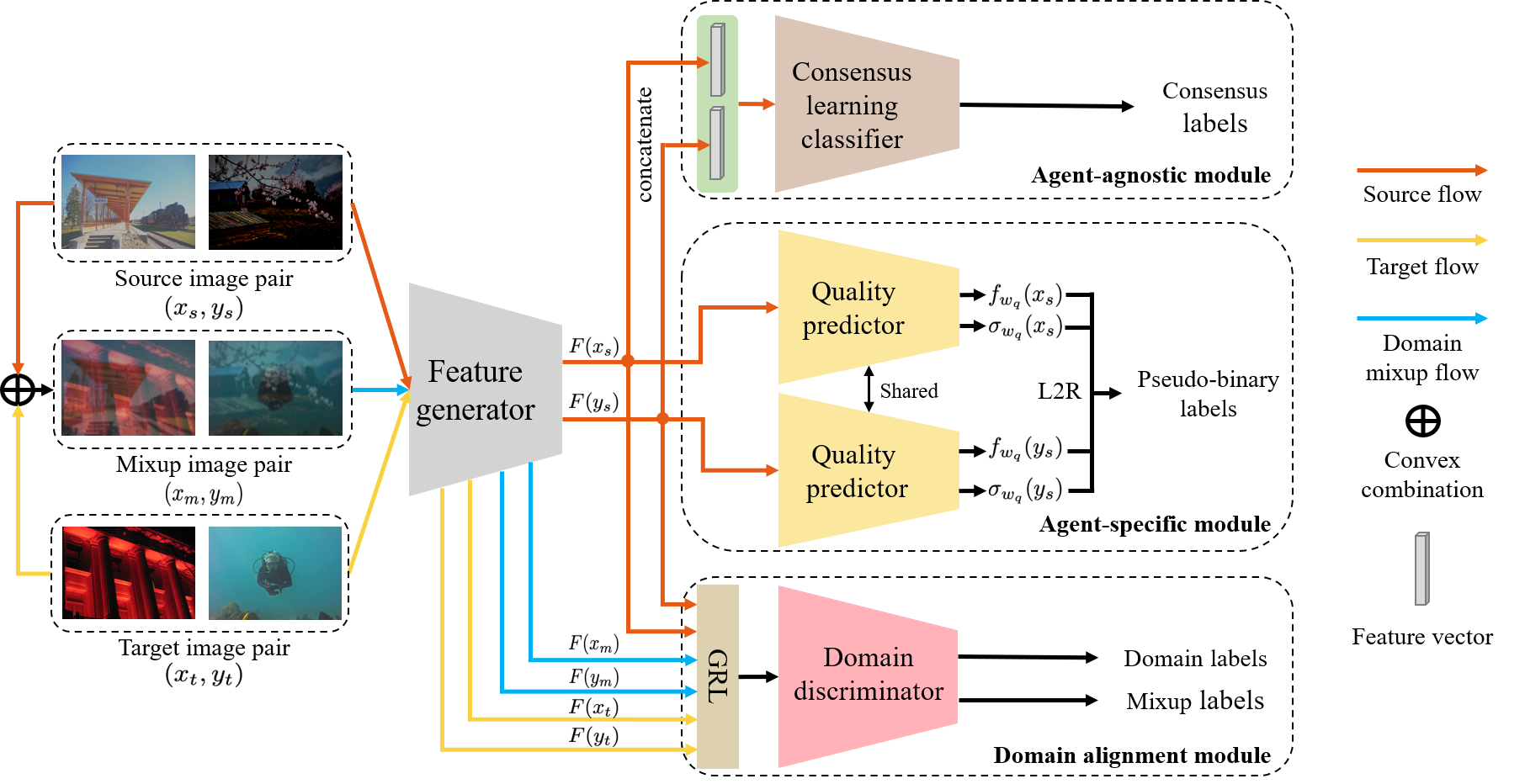}
	\end{center}
	\caption{The overall pipeline of $opinion$-$free$ BIQA model training. The quality predictor takes synthetic image pairs and their corresponding pseudo-binary labels as input to learn a perceptual quality prediction function $f_{w_q}$. The domain discriminator tries to reduce the domain shift between synthetic or authentic images. A consensus learning classifier (CLC) provides additional regularization for the rectification of feature extractor. The learned $f_{w_q}$ is used for the task of image quality predictions in the wild.}
	\label{fig:pipeline}
\end{figure*}

\subsection{Opinion-free BIQA}
\label{sec:pseudo-labels}
The annotation process for IQA image datasets requires multiple human scores for each image, and thus the acquisition procedure is extremely labor-intensive and costly (see TABLE \ref{tab:dataset}). Developing IQA models without human scores ($i.e.$, $opinion$-$free$ BIQA models) \cite{mittal2013making} is of great interest. One line for $opinion$-$free$ BIQA is making use of natural scene statistic (NSS) features. Mittal $et~al.$ \cite{mittal2011blind} proposed the first $opinion$-$free$ BIQA model - TMIQ, where probabilistic latent semantic analysis was applied to characterize quality-aware visual words based on NSS features between pristine and distorted images. Later, Mittal $et~al.$ \cite{mittal2013making} introduced NIQE - which is a "completely blind" image quality analyzer without access to any distorted images. NIQE fits the distribution of NSS feature vectors computed from a corpus of pristine image patches with a multivariate Gaussian model (MGV). A Bhattacharyya-like distance between the MVGs of high-quality images and target distorted images was computed as the quality score. Zhang $et~al.$ \cite{zhang2015feature} further extended NIQE to Integrated Local NIQE (ILNIQE) by incorporating more additional types of quality-aware NSS features and fitting more feature vectors (each patch per MVG) of target distorted image to compute local quality scores on them. They argued that local characteristics could increase the robustness of a BIQA model that is sensitive to local distortion artifacts.

The other line is to learn from pseudo-ground truth, $e.g.,$ pseudo-scores \cite{ye2014beyond}, pseudo-ranking order \cite{liu2017rankiqa}. Ye $et~al.$ \cite{ye2014beyond} proposed blind learning of image quality using synthetic scores (BLISS). Instead of being trained with human-opinion scores, BLISS was trained on synthetic scores derived from FR-IQA measures making use of reciprocal rank fusion (RRF). Wu $et~al.$ \cite{wu2020end} combined the quality values predicted by multiple FR-IQA models according to distortion types since the performance of different FR-IQA models on some specific distortion types may be different from each other. Liu $et~al.$ \cite{liu2017rankiqa} presented RankIQA. The supervised signals to train RankIQA is the rank order between images at different distortion levels but with the same distortion type,  thus lacking cross-distortion-type ranking information. Ma $et~al.$ \cite{ma2017dipiq} proposed dipIQ, where quality-discriminative image pairs were automatically obtained and labeled by three-trusted FR-IQA models, $i.e.,$ MS-SSIM, VIF, and GSMD, achieving excellent performance on synthetically-distorted IQA datasets. To sum up, progress has been made on developing $opinion$-$free$ BIQA models, but mainly on \textit{synthetic} IQA scenarios. Few of them can perform satisfactory quality assessment on \textit{authentic} images captured in the \textit{wild}.

\section{Method}
% In this section, we detail our proposed methods. 
Our goal is to design a CNN-based BIQA model that is able to learn from an unlabeled synthetically-distorted dataset (source domain) $\mathcal{X}_s$ = $\{x_s^{(i)}\}_{i=1}^{N_s}$ without using human opinion scores that is of high performance to assess the image quality of authentically-distorted images (target domain) $\mathcal{X}_t$ = $\{x_t^{(j)}\}_{j=1}^{N_t}$ captured in the wild. Here, $N_s$ and $N_t$ are the number of images in $\mathcal{X}_s$ and $\mathcal{X}_t$ respectively. %In many scenarios, precise ground-truths can be obtained during the creation of a simulated dataset \cite{mu2020learning, varol2017learning}. Liu. $et~al.$ exploited this strategy where the different severities of human-made distortion with the same distortion types are treated as supervisory signals to train the DNN-based BIQA model. However, the cross-distortion-type labels are not able to obtain, leading to poor performance if not fine-tuned on the labeled dataset. 
Fig. \ref{fig:pipeline} depicts the overall structure of our method. Our unified model contains three modules: agent-specific module, agent-agnostic module, and domain alignment module. In our design, the supervisory signals are from a set of high-performance FR-models \cite{ye2014beyond, ma2019blind, wu2020end}, termed as \textit{agent} in our context since there are no human opinion scores. It was noted that when assessing the visual quality of an image, those agents agree to disagree, i.e.,  they have both \textit{consensus} and \textit{specific} information. Thus we introduce an agent-specific and an agent-agnostic module to respectively capture the consensus and specific information from the annotations of a set of high-performance agents of FR-models.  % Motivated by the works learning from multiple annotators \cite{ye2014beyond, ma2019blind, wu2020end}, the supervisory signals are annotated by a set of high-performance FR-models.
The consensus information is captured by a consensus learning classifier (CLC), and module-specific information is implicitly learned by forcing the model to mimic the performance of each agent. %\textbf{(it seems that we need to have a residual layer here to minus the consensus information.)}. 
Due to the existence of domain shift between synthetically-distorted images and authentically-distorted ones, we further design a domain \textit{alignment} module with a domain discriminator to alleviate this issue.  As shown in Fig. \ref{fig:pipeline}, we denote the feature generator, quality predictor, CLC, domain discriminator as $F$, $Q$, $C$, $D$, respectively. Since the parameters $w_f$ of F are shared among the modules $Q$, $C$, $D$, for the sake of clarity, we denoted the union of the parameters of $F$ and $Q$ as $w_q$, the union of the parameters of $F$ and $C$ as $w_c$, and the union of the parameters of $F$ and $D$ as $w_d$, respectively. Then the learned BIQA model can be defined as $f_{w_q}$.
    
\subsection{Modules for Quality Prediction}
\noindent
\textbf{Agent-specific module:} Given an image $x_s \in \mathcal{X}_s$, let $f(x_s)$ represent its true perceptual quality. We utilize $M$ full-reference IQA models $\{f^m\}_{m=1}^{M}$ ($i.e.,$ $M$ agents) to estimate  $f(x_s)$, which are collectively denoted as $\{f^m(x_s)\}_{m=1}^{M}$. It was noted that different agents have different range of predictions, calibrating their outputs will be difficult and unnecessary. Therefore, directly using the output of each agent does not suit directly. To cope with this issue, for an image pair, instead of directy usng the predictions of the agents as pseudo labels, we used a rank label inferred based on the outputs by a single agent. In this way, the leaning will focus on the rank of perceptual qualities, rather than an absolute value of the visual quality. Thus for one image pair, in total, we will have $M$ pseudo-labels, each per agent. Specifically, we first randomly sample $N$ image pairs $\{(x_s^{(i)}, y_s^{(i)})\}_{i=1}^{N}$ from $\mathcal{X}_s$. For each image pair $(x_s, y_s)$, let $q_m$ denote its pseudo binary ranking label, with \textit{one} (\textit{zero}) indicating the perceptual quality of $x_s$ is ranked  \textit{higher} (\textit{lower})  than $y_s$ by $m$-th agent model, $i.e.$,  $q_m$ = $1$ (that is $positive$) if $f^m(x_s) \ge  f^m(y_s)$ and $q_m$ = $0$ (that is $negative$) otherwise. In this way, we are able to build a training set of image pairs $\mathcal{D}=\{(x_s^{(i)}, y_s^{(i)}), q_1^{(i)}, \cdots, q_m^{(i)}\}_{i=1}^{N}$, where $x_s, y_s \in \mathcal{X}_s$.
    
Since the quality estimates of each agent are noisy and not equally good (or bad) at labeling the input, we introduce two uncertainty variables, $i.e.,$ the probabilities of hit rate $\alpha_m$ and correct reject rate $\beta_m$ to model the reliability of $m$-th agent of IQA annotator, which are denoted in the form of conditional probabilities as:
     \begin{align}
       \alpha_m= \Pr(q_m = 1 | q = 1)
       \label{eq:alpha}
    \end{align}
    and
    \begin{align}
      \beta_m= \Pr(q_m = 0 | q = 0),
      \label{eq:beta}
    \end{align}
    respectively, where $q=1$ if $f(x_s)\ge f(y_s)$ and $q=0$ otherwise.
    
\begin{figure*}[t]
	\centering
	% \captionsetup{justification=centering}
	\subfloat[Blur]{\includegraphics[width=0.19\textwidth]{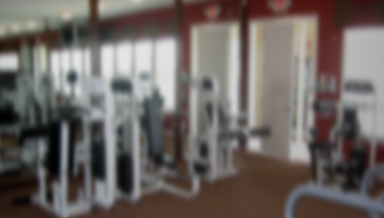}}\hskip.25em
	\subfloat[Motion blur]{\includegraphics[width=0.19\textwidth]{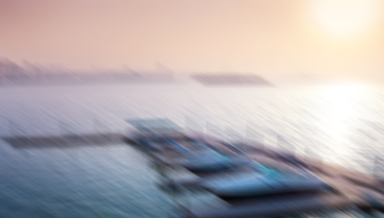}}\hskip.25em
	\subfloat[Noise]{\includegraphics[width=0.19\textwidth]{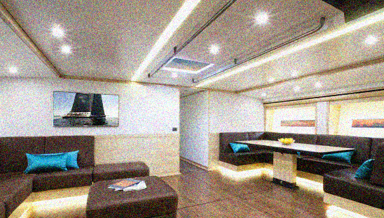}}\hskip.25em
	\subfloat[Overexposure]{\includegraphics[width=0.19\textwidth]{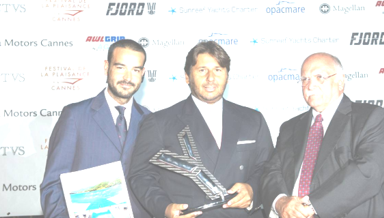}}\hskip.25em
	\subfloat[Underexposure]{\includegraphics[width=0.19\textwidth]{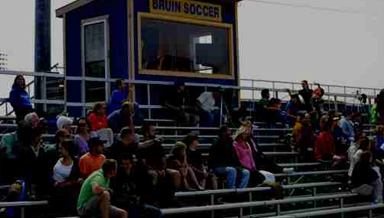}}\\
	\subfloat[Blur]{\includegraphics[width=0.19\textwidth]{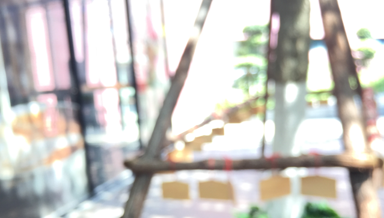}}\hskip.25em
	\subfloat[Motion blur]{\includegraphics[width=0.19\textwidth]{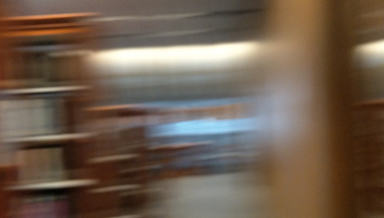}}\hskip.25em
  	\subfloat[Noise]{\includegraphics[width=0.19\textwidth]{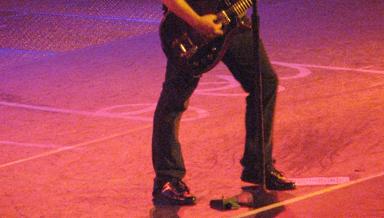}}\hskip.25em
	\subfloat[Overexposure]{\includegraphics[width=0.19\textwidth]{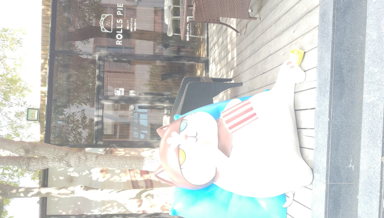}}\hskip.25em
	\subfloat[Underexposure]{\includegraphics[width=0.19\textwidth]{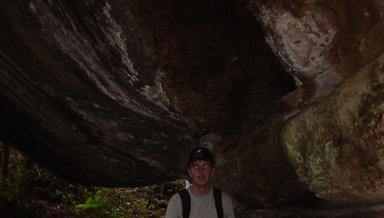}}\\
	\caption{Representative samples of source domain - synthetically-distorted dataset $\mathcal{X}_s$ which are generated from Waterloo Exploration Database \cite{ma2016waterloo} with man-made distortions, and target domain which are from two authentically-distorted datasets - SPAQ \cite{fang2020cvpr} or KonIQ-10k \cite{hosu2020koniq}. \textbf{(a)}-\textbf{(e)} are synthetically-distorted images from $\mathcal{X}_s$. \textbf{(f)}-\textbf{(j)} are authentically-distorted images from SPAQ or KonIQ-10k, with the same dominant distortion type. For these samples, we can intuitively observe the proximity of the synthetic and authentically distortions. 
	%Images in the right panel has more diversity then those on the left panel.
% 	In this setting, the selected gMAD images have limited diversity, most of which are underexposure or striped images. 
% 	The predicted values are mapped onto the CLIVE MOS scale, with a higher number indicating better perceptual quality. 
	}
	\label{fig:syn_aut}
\end{figure*}

    Under the hypothesis of Thurstone’s Case V model \cite{thurstone1927law}, we make the assumption that the true perceptual quality $f(x)$ obeys a Gaussian distribution with mean and variance estimated by $f_{w_q}(x)$ and $\sigma_{w_q}(x)$, where $w_q$ is the model parameter we want to learn. That is to say, $f_{w_q}(\cdot)$ and $\sigma_{w_q}(\cdot)$ are two differentiable functions we want to learn for image quality prediction and the corresponding standard deviation (std) (uncertainty) estimation, respectively. Then, the probability of preferring $x_s$ over $y_s$ perceptually in an image pair $(x_s, y_s)$ can be computed from the Gaussian cumulative distribution function $\Phi(\cdot)$, which has a close-form as 
    \begin{align}
    	p_{w_q}(x_s,y_s)&= 
    	\Phi\left(\frac{f_{{w_q}}({x_s}) - f_{{w_q}}({y_s})}{\sqrt{\sigma_{w_q}^2(x_s)+\sigma_{w_q}^2(y_s)}}\right).
    	\label{eq:cdf}
    \end{align}
    During training, the model parameters $w_q$ along with the uncertainty variables $\{\alpha, \beta\}$ = $\{\alpha_m, \beta_m\}_{m=1}^M$ are jointly optimized using maximum likelihood function~\cite{ma2019blind}, which is defined as follows: 
    \begin{align} 
    	\{\hat{w_q}, \hat{\alpha},\hat{\beta}\} = \mathop{\text{argmax}}_{w_q,\alpha,\beta} \Pr(\mathcal{D};w_q,\alpha,\beta),
    	\label{eq:ml19}
    \end{align}
    where 
    \begin{align}
    	\Pr(\mathcal{D};w_q,\alpha,\beta) =& \prod_{i=1}^N\bigg(p_{w_q}(x_s^{(i)},y_s^{(i)})\prod_{m=1}^M\Pr(q_m^{(i)}|q=1)\nonumber\\
    	+(1-&p_{w_q}(x_s^{(i)},y_s^{(i)}))\prod_{m=1}^M\Pr(q_m^{(i)}|q=0)\bigg).   
    \end{align}
    In practice, we minimize the negative logarithm version of Eq. (\ref{eq:ml19}). Therefore, the optimization objective over a min-batch  $\mathcal{B}_s \subset\mathcal{D}$ is equivalently as:
     \begin{align} 
     \begin{split}
    	\ell_q(\mathcal{B}_s; w_q, \alpha, \beta) =& \\
    	-\frac{1}{\left|\mathcal{B}_s\right|}\sum_{i=1}^{\left|\mathcal{B}_s\right|} \log\Pr&(x_s^{(i)},y_s^{(i)}, \{q^{(i)}_m\}_{m=1}^M ;w_q,\alpha,\beta),
    \end{split}
    \label{eq:lq}
    \end{align}
    where $\left|\mathcal{B}_s\right|$ denotes the cardinality of $\mathcal{B}_s$.
    
    \begin{figure}[t]
	\begin{center}
		\includegraphics[width=0.8\linewidth]{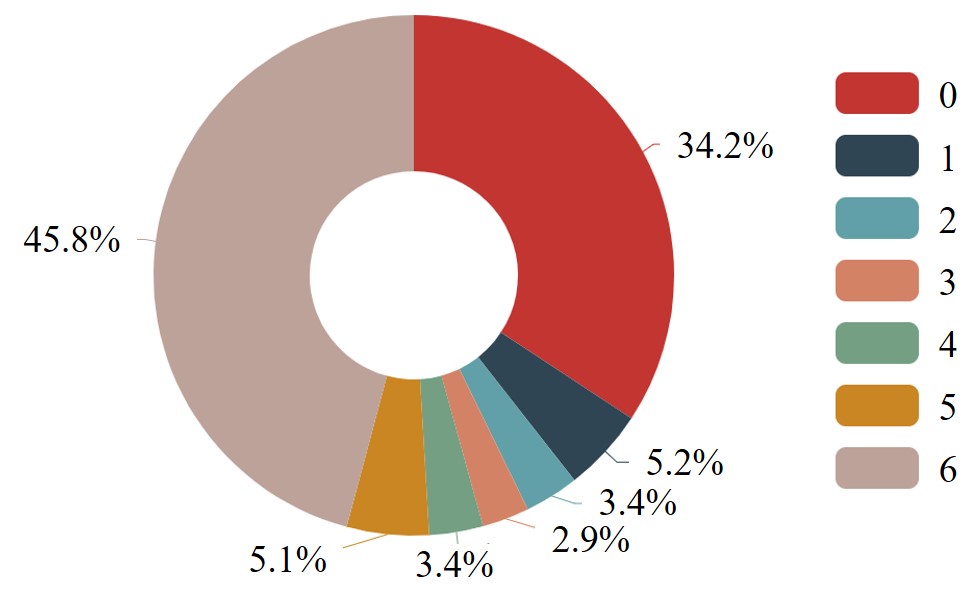}
	\end{center}
	\caption{Percentage of image pairs labeled as \textit{positive} by six FR-IQA agents in $\mathcal{D}$. For example, there are $5.1$\% of image pairs having \textit{five} positive labels, and $34.2$\% of image pairs having \textit{no} positive labels, etc. In total, roughly $80$\% of image pairs are in completely agreement with six FR-IQA models (i.e., each pair having six labels all as \textit{positive} or all as \textit{negative}).}
	\label{fig:distribution}
    \end{figure}
    \vspace{0.3em}
    \noindent
    \textbf{Agent-agnostic module:} To learn the consensus information among the multiple agent annotators, we further add a consensus learning classifier  $C$ as a complementary objective in our pipeline, which can boost final model performance. In another view, the complementary objective could also be considered as a diverse task from quality prediction but highly related to it \cite{chen2020adversarial, pang2019improving}. Learning the agreement of different agents could improve the performance of an underlying model \cite{penrose1946elementary, lam1997application}. 
    %a CLC is motivated with the ensemble of self-supervised tasks for representation learning \cite{chen2020adversarial}. \textbf{JZ: the following statement doesn't support the design of CLC, please rewrite. (Strictly, our learning pipeline is also a line of the self-supervised pre-training framework. Many works \cite{chen2020adversarial, pang2019improving} have demonstrated that diversity can challenge the transfer-ability of improved ensemble performance. Inspired by this, we design the additional task as a classification task to promote ensemble diversity instead of another quality prediction task.)} 
    Inspired by this, the supplementary objective is devised as a consensus classification task with the aim of refining feature extractor $F$ to promote the performance of quality prediction.
    Technically, the input to CLC is the concatenated feature embeddings of an image pair ($x_s$, $y_s$) $\in~\mathcal{D}$, $i.e.,$ the concatenation of $F(x_s)$ and $F(y_s)$ (see Fig. \ref{fig:pipeline} for illustration). The supervising signal is generated by the application of majority voting of the pseudo labels, $i.e.,$ $l_c = 1$ if $\sum_{m=1}^M q_m \geqslant M/2$ else $l_c = 0$ otherwise. Other combination implementations, $e.g.,$ Bayesian, logistic regression, fuzzy integral and neural network \cite{lam1997application} may also be plausible. The loss function of CLC for a mini-batch $\mathcal{B}_s$ sampled from $\mathcal{D}$ is defined as follows:
    \begin{align}
        \begin{split}
            \ell_c(\mathcal{B}_s; w_c) & =  -\frac{1}{\left|\mathcal{B}_s\right|}\sum_{i=1}^{\left|\mathcal{B}_s\right|}\bigg(l_cg(C(F(x_s^{(i)}) \oplus F(y_s^{(i)})))\\
            & +(1-l_c)g(1-C(F(x_s^{(i)}) \oplus F(y_s^{(i)})))\bigg),
       \end{split}
    \end{align}
    %\end{align}
    % \begin{align}
    %   	\ell_v&(x_s,  y_s; w_v) = \\\nonumber -& l_v(1-V(F(x_s) \oplus F(y_s)))^{\gamma}\log (V(F(x_s) \oplus F(y_s))) \\\nonumber
    %   	-& (1-l_v)(V(F(x_s) \oplus F(y_s)))^{\gamma}\log (1-V(F(x_s) \oplus F(y_s))),
    %   	%	\label{eq:tloss}
    % \end{align}
    where $\oplus$ is the feature vector concatenation operation. $g(x)=(1-x)^\gamma\log (x)$ is an adaptive loss function (ADL) that can assign high weights on hard-to-classify samples (see Section \ref{para:focal_loss} for detailed definition). Other types of feature vector fusion, $e.g.,$ the residual $F(x_s^{(i)}) - F(y_s^{(i)})$, the concatenation of feature vectors and their residual $F(x_s^{(i)}) \oplus F(y_s^{(i)}) \oplus (F(x_s^{(i)}) - F(y_s^{(i)}))$ \cite{bosse2017deep}, are also suitable in our setting.  
    
\subsection{Modules for Domain Alignment}
\label{para:focal_loss}
	In the wild, images are easily distorted by poor lighting conditions, sensor limitations, lens imperfections, amateur manipulations, and etc. \cite{zhang2019learning}. These distortions are very complex and almost impossible to synthesize in the laboratory \cite{ghadiyaram2015massive}. This may cause BIQA models learned only from synthetically-distorted images $\mathcal{D}$ (source domain) to generalize poorly on authentically-distorted images $\mathcal{X}_t$ (target domain) \cite{ganin2016domain, tzeng2017adversarial, xu2020adversarial}. To alleviate this issue, we design a module for domain alignmentb based UDA to accommodate the features learned from synthetic images to authentic images. This module contains two components: 1) adversarial domain alignment and 2) domain mixup. 
	
    \begin{algorithm}[t] 
    	\SetAlgoLined
    	\KwIn{A synthetically-distorted image set $\mathcal{X}_s$ as source domain, an authentically-distorted image set $\mathcal{X}_t$ as target domain, $M$ full-reference IQA models  $\{q_m\}_{m=1}^{M}$, trade-off hyper-parameters $\lambda_1$, $\lambda_2$ and $\lambda_3$, maximum epochs $T_1$ for merely training $f_{w_q}$, maximum epochs $T_2$ for domain confusion}
    	
    	\KwOut{An optimized BIQA model $f_{w_q}$}
    	Pair-wisely sample $n$ image pairs $\mathcal \{(x_s^{(i)},y_s^{(i)})\}_{i=1}^{N}$ from $\mathcal{X}_s$ \\
    	Compute the binary labels $\{q_m^{(i)}\}_{m=1}^M$ for each pair in $\{(x^{(i)},y^{(i)})\}_{i=1}^{N}$ to create $\mathcal{D}$ = $\{(x_s^{(i)}, y_s^{(i)}), q_1^{(i)}, \cdots, q_M^{(i)}\}_{i=1}^{N}$\\    	
    	\For{$t \gets~1~\KwTo~T_1$}{
    		Train $f_{w_q}$ by minimizing Eq.~(\ref{eq:lq}) on $\mathcal{D}$
    	}
    	\For{$t \gets~T_1+1~\KwTo~T_1+T_2$}{
    	Adapt $f_{w_q}$ to $\mathcal{X}_t$ by minimizing Eq.~(\ref{eq:tloss}) on $\mathcal{D}$ and $\mathcal{X}_t$ 
        } 
    	\caption{Learning from the synthetic for BIQA in the wild}
    	\label{al:algorithm1}
    \end{algorithm}
    
\begin{figure*}[t]
	\begin{center}
		\includegraphics[width=1.0\linewidth]{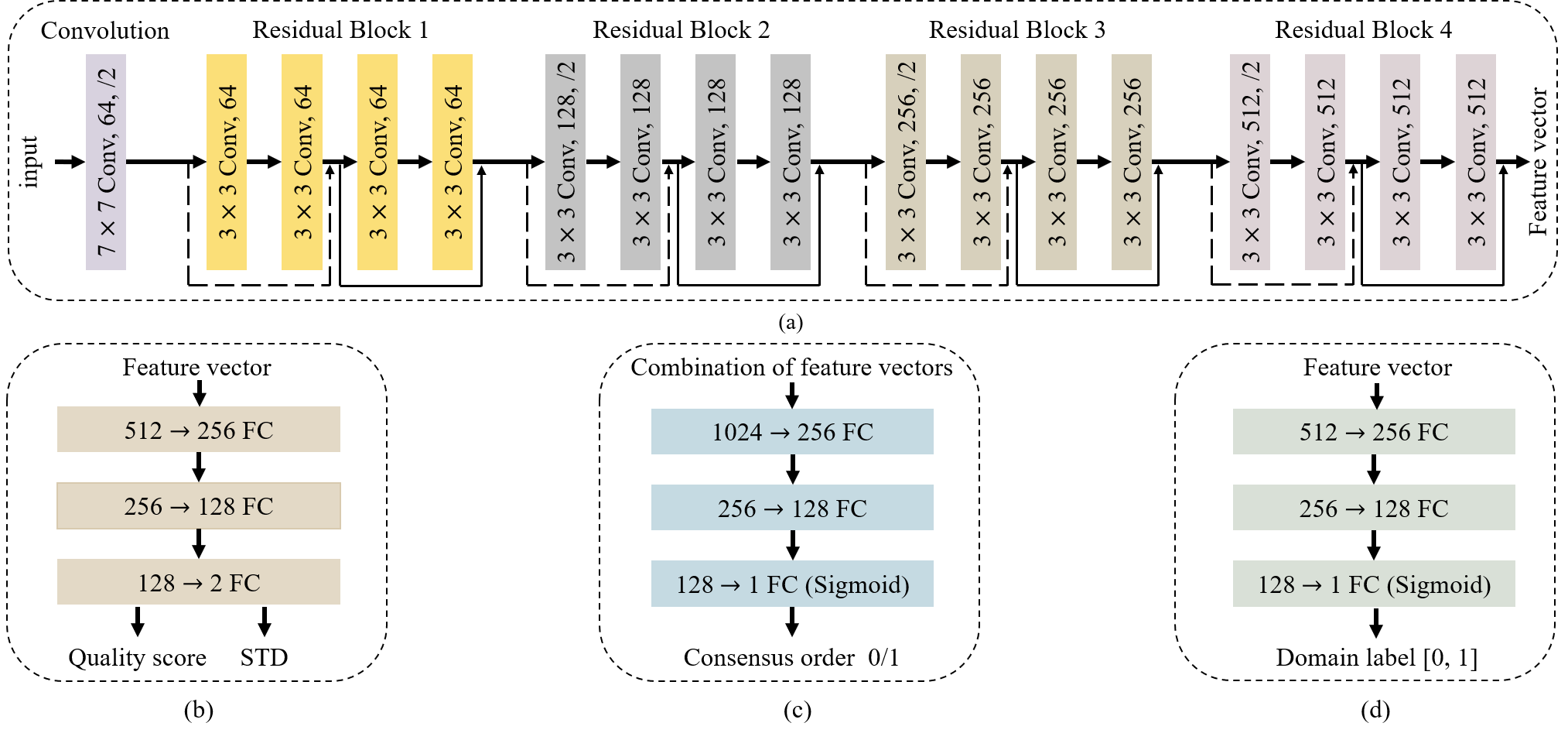}
	\end{center}
	\caption{The detailed architecture of network components with input and output dimensions. \textbf{(a)} Feature extractor $F$, where the dotted shortcuts increase dimensions. \textbf{(b)} Quality predictor $Q$, where the outputs are quality score and corresponding STD of image. \textbf{(c)} Consensus learning classifier (CLC) $C$, where input is combination of feature vectors of image pair. \textbf{(d)} Domain discriminator $D$. }
	\label{fig:network}
\end{figure*}
    
    \noindent
    \textbf{Adversarial domain alignment:} Inspired by the works of generative adversarial learning (GAN) \cite{goodfellow2014generative} and DANN \cite{ganin2016domain}, we introduce an additional domain discriminator $D$ to distinguish whether an input $x$ is from the source domain $\mathcal{X}_s$ or the target domain $\mathcal{X}_t$ (see Fig. \ref{fig:pipeline} for details). The feature generator $F$ is trained to produce images in a way that confuses the discriminator, which in turn ensures that the network cannot distinguish between the distributions of $\mathcal{X}_s$ and $\mathcal{X}_t$.  Given two samples $x_s \in \mathcal{X}_s$ with manually assigned domain label $l_s = 1$ and $x_t \in \mathcal{X}_t$ with assigned domain label $l_t = 0$, the objective function to learn the domain-invariant features is formulated as follows by minimizing the cross-entropy loss: 
    \begin{align}
    \begin{split}
       	\ell_d(\mathcal{X}_s,\mathcal{X}_t; w_d) =& -\sum_{i=1}^{N_s}\log (D(F(x_s^{(i)})))\\-& \sum_{j=1}^{N_t}\log (1-D(F(x_t^{(j)}))).
    \end{split}
    \label{eq:celoss}
    \end{align}
    It was aware that for the samples with small domain shifts, models without UDA might work well, while the transfer performance is likely to be hurt by the samples with large shifts \cite{saito2018strong}. Therefore, ideally, cross-entropy should assign a non-negligible loss to these easy-to-classify examples (samples with small domain shift) and large ones to hard-to-classify samples (i.e., samples with large domain shifts). To achieve this goal, an ADL is introduced to put larger weights on hard samples than on those easy ones during training. Therefore, we reformulate Eq. (\ref{eq:celoss}) over a mini-batch
    %\textbf{(JZ:It seems that there are some issues of the labelling of equations. Please double check.)} 
    as follows:
    \begin{align}
    \begin{split}
        \ell_d(\mathcal{B}_s,\mathcal{B}_t; w_d) = &-\frac{1}{\left|\mathcal{B}_s\right|}\sum_{i=1}^{\left|\mathcal{B}_s\right|} g(D(F(x_s^{(i)})))\\
        &-\frac{1}{\left|\mathcal{B}_t\right|}\sum_{j=1}^{\left|\mathcal{B}_t\right|}g(1-D(F(x_t^{(j)}))),
    \end{split}
    \label{eq:flloss}
    \end{align}
    % \begin{align}
    %   	\ell_d(\mathcal{X}_s,&\mathcal{X}_t; w_d) = -\sum_{i=1}^{N_s} (1-D(F(x_s^{(i)})))^{\gamma}\log (D(F(x_s^{(i)}))) \\\nonumber
    %   	-&\sum_{j=1}^{N_t}(D(F(x_t^{(j)})))^{\gamma}\log (1-D(F(x_t^{(j)}))),
    %   	%\label{eq:flloss}
    % \end{align}
    where $\mathcal{B}_s$ and $\mathcal{B}_t$ are two min-batches sampled from $\mathcal{D}$ and $\mathcal{X}_t$, respectively. $g(x) = (1-x)^\gamma\log(x)$ is an ADL applied on $x$, and $\gamma$ controls the weight applied on the samples with large domain shift. Between the feature extractor and domain discriminator, we insert a \textit{gradient reversal layer} (GRL) \cite{ganin2016domain} in the architecture to facilitate the optimization. In the forward pass of the network, the GRL acts as an identity function while reverses the sign of gradients during the backpropagation. In this way, after training, it is hard for the domain discriminator to judge whether the feature of a sample is from the source or target domain ($i.e.$, samples' domain labels are as indistinguishable as possible for the domain classifier), thus resulting in the domain-invariant features. % \cite{ganin2016domain}

    \noindent
    \textbf{Domain mixup:} To further enhance the generalizability and robustness of adaptation models, pixel-level domain mixup \cite{xu2020adversarial} is employed to drive the domain discriminator to explore the inter-domain information of two distributions. Let $x_s \in \mathcal{D}$ and $x_t \in \mathcal{X}_t$ represent two samples from each domain, and the pixel-level mixup is linearly interpolated to produce a mixup image $x_m$ and corresponding soft domain labels $l_{m}$: 
    \begin{align}
       	x_m = \lambda x_s + (1 - \lambda)x_t, \\ 
       	l_m = \lambda l_s + (1 - \lambda) l_t = \lambda,
       	%\label{eq:flloss}
    \end{align}
    where $\lambda \in$ [$0$, $1$] is the mixup ratio, and $\lambda \thicksim$ Beta($\xi$, $\xi$), in which $\xi$ is set to a constant $2.0$ in all experiments. The mixup loss of $x_m$ is defined as follows:
    \begin{align}
    \begin{split}
      	\ell_m(\mathcal{B}_m; w_d) = -& \frac{1}{\left|\mathcal{B}_m\right|}\sum_{i=1}^{\left|\mathcal{B}_m\right|}\bigg( l_mg(D(F(x^{(i)}_m)))\\
      	+&(1-l_m)g(1-D(F(x^{(i)}_m)))\bigg),
    \end{split}
    \label{eq:lmloss}
    \end{align}
    % \begin{align}
    %   	\ell_m(\mathcal{X}_s,\mathcal{X}_t; w_d) = -& \sum_{i=1}^{N_m} l_m(1-D(F(x^{(i)}_m)))^{\gamma}\log (D(F(x^{(i)}_m))) \\\nonumber
    %   	-\sum_{i=1}^{N_m}(1-l_m)&(D(F(x^{(i)}_m)))^{\gamma}\log (1-D(F(x^{(i)}_m))),
    %   	%\label{eq:lmloss}
    % \end{align}
    where $\mathcal{B}_m$ is a mixup mini-batch of $\mathcal{B}_s$ and $\mathcal{B}_t$, and $g(\cdot)$ is the ADL function. Through the introduction of pixel-level mixup, the feature generator can refine the robustness and generalization ability of $f_{w_q}$, especially when there exists distribution oscillation in the test phase \cite{xu2020adversarial}.
    
    \subsection{Joint Learning and Adaptation}
    In the training phase, the final joint optimization objective function can be written as:
    \begin{align}
    \begin{split}
       	\ell(\mathcal{B}_s, \mathcal{B}_t, \mathcal{B}_m) &= \ell_q(\mathcal{B}_s) + \lambda_1 \ell_c(\mathcal{B}_s) \\
       	   & + \lambda_2 \ell_d(\mathcal{B}_s, \mathcal{B}_t) + \lambda_3\ell_m(\mathcal{B}_m),
    \end{split}
    \label{eq:tloss}
    \end{align}
    where $\lambda_1$, $\lambda_2$ and $\lambda_3$ are hyperparameters that control the trade-offs between quality assessment learning and domain adaptation. Stochastic gradient descent (SGD) is used to update the network parameter vectors $\{w_q, w_d, w_c, \alpha, \beta\}$. In the test phase, the learned feature extractor and quality predictor are used for the final assessment of input images. 
    %\textbf{(JZ: I would expect we need to discuss how the values of those parameters controling the tradeoff are set.)}
    % \begin{align}
    %   	\ell(\mathcal{B}_{\mathcal{D}},  & \mathcal{B}_t; w_q,\alpha,\beta,w_d)\nonumber\\ 
    %   	&= \frac{1}{|\mathcal{B}_\mathcal{D}|} \sum_{i=1}^{|\mathcal{B}_\mathcal{D}|} 	\ell_q(x_s^{(i)},y_s^{(i)}, \{q_m^{(i)}\}_{m=1}^M; w_q, \alpha, \beta)\nonumber \\ & + \lambda_1 \frac{1}{|\mathcal{B}_t|} \sum_{j=1}^{|\mathcal{B}_t|} \ell_d(x_s^{(j)}, x_t^{(j)}; w_d) \\\nonumber + 
    %   	\label{eq:tloss}
    % \end{align}
    % where  $|\mathcal{B}_\mathcal{D}|$  and  $|\mathcal{B}_t|$ represent the cardinality of $\mathcal{B}_\mathcal{D}$ and $\mathcal{B}_t$, respectively, and $\lambda$ is a hyper-parameter to trade off two terms. Algorithm \ref{al:algorithm1} summarizes the whole procedure of the proposed method.

%\IEEEpubidadjcol

\section{Experiment and Results}
To demonstrate the efficacy of our method, a set of experiments are conducted. We first describe the experimental details and then present the extensive experiment results. Our results show that our proposed method can learn a state-of-the-art $opinion$-$free$ BIQA model for image quality prediction in the wild.  %We also show that a large number of unlabeled synthetic data can benefit the training of BIQA models in the self-supervised proxy task.

 \begin{table}[t]
    \caption{Correlations between model predictions and MOSs on LIVE \cite{sheikh2006statistical} (a synthetic IQA dataset as reference), KonIQ-10k~\cite{hosu2020koniq} and SPAQ \cite{fang2020cvpr}, respectively. In each sub-table, top, middle, and bottom sections list performances of three knowledge-driven, three DNN-based, and our proposed  $opinion$-$free$ BIQA models, respectively. Ours($\mathcal{X}_t$) with different $\mathcal{X}_t$ indicates our methods being applied to different target domain, and \textit{na\"{i}ve} means no target domain. The top two correlations obtained by BIQA models are highlighted in boldface}
    \label{table:comparision}
    \vspace{-.3cm}
	\begin{center}
	    \begin{threeparttable}
		\begin{tabular}{l|ccc}
    		\toprule[1pt]
			SRCC & LIVE\cite{sheikh2006statistical}&KonIQ-10k\cite{hosu2020koniq}& SPAQ\cite{fang2020cvpr}\\
			\hline
			QAC \cite{xue2013learning}& $0.868$ & $0.092$ & $0.340$\\
			NIQE \cite{mittal2013making}& $0.906$ & $0.530$ & $0.703$ \\
 			ILNIQE \cite{zhang2015feature}& $0.898$ & $0.506$ & $0.714$ \\
 			\hline
 			RankIQA\tnote{1} \cite{liu2017rankiqa}& $0.898$ & $0.483$ & $0.584$\\
 			dipIQ \cite{ma2017dipiq} & $\textbf{0.938}$ & $0.236$ & $0.385$ \\
 			Ma19 \cite{ma2019blind} & $0.919$ & $0.456$ & $0.379$ \\
			\hline
			Ours(na\"{i}ve) & $0.856$ & $0.634$ & $0.807$\\
			Ours(KonIQ-10k) & $0.907$ & $\textbf{0.717}$ & $\textbf{0.826}$ \\
			Ours(SPAQ) & $\textbf{0.920}$ & $\textbf{0.712}$ & $\textbf{0.838}$ \\
			\hline
			\hline
			PLCC & LIVE & KonIQ-10k  & SPAQ \\
		    \hline
			QAC  & $0.863$ & $0.244$ & $0.371$\\
			NIQE  & $0.904$ & $0.538$ & $0.712$ \\
 			ILNIQE  & $0.903$ & $0.531$ & $0.721$ \\
 			\hline
 			RankIQA & $0.892$ & $0.482$ & $0.587$\\
 			dipIQ & $\textbf{0.935}$ & $0.435$ & $0.497$ \\
 			Ma19 & $\textbf{0.917}$ & $0.462$ & $0.391$ \\
			\hline
			Ours(na\"{i}ve) & $0.849$ & $0.652$ & $0.813$\\
			Ours(KonIQ-10k) & $0.906$ & $\textbf{0.740}$ & $\textbf{0.831}$ \\
			Ours(SPAQ) & $0.912$ & $\textbf{0.736}$ & $\textbf{0.844}$ \\
			\bottomrule[1pt]
		\end{tabular}
		\begin{tablenotes}
             \item[1] https://github.com/YunanZhu/Pytorch-TestRankIQA.
         \end{tablenotes}
    \end{threeparttable}
	\end{center}
\end{table}

\subsection{Datasets}
Four datasets are used in our experiment to test the performance of the proposed method, including Waterloo Exploration Database \cite{ma2016waterloo} to construct $\mathcal{X}_s$  as the source domain, two large-scale authentically-distorted datasets - KonIQ-10k \cite{hosu2020koniq} and SPAQ \cite{fang2020cvpr} behaving as $\mathcal{X}_t$ for domain adaptation and performance evaluation, and a newly established authentically-distorted dataset, including more than $100K$ images, as gMAD playground for the test of generalizability.

\vspace{.3em}
\noindent\textbf{Waterloo Exploration Dataset \cite{ma2016waterloo}}: This dataset consists of $4,744$ high-quality natural images collected from the Internet, with a large span of visual content diversity. It was mainly used for testing the generalizability of IQA models, $e.g.,$ P-test, L-test, D-test, as well as gMAD competition \cite{ma2016waterloo, ma2018group}. Liu $et~al.$ \cite{liu2017rankiqa} and Ma $et~al.$ \cite{ma2019blind} simulated synthetically-distorted datasets based on the pristine images to train CNN-based BIQA models. Here, we also make use of the Waterloo dataset for generating $\mathcal{X}_s$. 

\vspace{.3em}
\noindent\textbf{KonIQ-10K \cite{hosu2020koniq}}: It is one of the largest IQA datasets consisting of $10,073$ quality scored images in the wild selected from $10$ million YFCC100m entries, presenting a broad range of diverse contents and authentic distortions. A total of $1.2$ million of reliable human ratings were obtained through crowdsourcing with $1,459$ crowd workers. As a result, each image has approximately $120$ quality scores.

\noindent\textbf{SPAQ \cite{fang2020cvpr}}: This dataset was constructed by Fang $et~al.$,  the largest to date smartphone photograph dataset (as named SPAQ). It consists of $11,125$ images captured by $66$ smartphones. Most of the images were captured by unprofessional photographers, where the capturing processes are easily affected by realistic camera distortions, $e.g.,$ lighting conditions, sensor limitations, lens imperfections. These artifacts are hard to be simulated by synthetic distortions. The human ratings are collected from a large subjective experiment conducted in a laboratory environment, where each image has about $15$ human ratings.

\noindent\textbf{Dataset for gMAD in the wild}: In order to test the generalizability of our proposed methods, we further collect a \textit{large-scale} \textit{authentically}-distorted image dataset\footnote{https://mega.nz/folder/G5ox0AoD\#QCykopZV1De6NSAEO334gQ} as the candidate pool for gMAD competition. Specifically, we first downloaded $750K$ images from Flicker\footnote{https://www.flickr.com/photos/tags/flicker/} with diverse real-world distortions. The near-duplicate images were then removed by the command-line tool $\mathrm{imgdupes}()$\footnote{https://github.com/knjcode/imgdupes\#against-large-dataset}, and those images of pure texts were also deleted, such as ppt, book pages, using OCR tool (pytesseract\footnote{https://pypi.org/project/pytesseract/}), and the images without Image Maker ($e.g.,$ Canon, Nikon, etc.) were de-duplicated. This resulted in a dataset of $635K$ naturally distorted photographic images. Afterwards, following the protocol in \cite{vonikakis2016shaping}, we sampled $100K$ images uniformly  $w.r.t$, image  attribute scores, $i.e.,$ bit-rate, JPEG compression ratio, brightness, colorfulness, contrast, and sharpness. Finally, the images were down-sampled to a maximum width or height of $1,024$ as a way to further reduce possible visible artifacts. %\textbf{(JZ: why down-sampling could reduce the artifacts? the reason is not clear here!)}. 
It is worth noting that this new dataset presents a \textit{wide} range of realistic distortions, such as sensor noise contamination, blurring (out-of-focus, motion), underexposure, overexposure, contrast reduction, color deviation, and a mixture of multiple distortions above, thus serving a good platform to deploy the gMAD competition to test the generalizability of our models.

\begin{figure*}[t]
	\centering
	% \captionsetup{justification=centering}
	\subfloat[]{\includegraphics[width=0.23\textwidth]{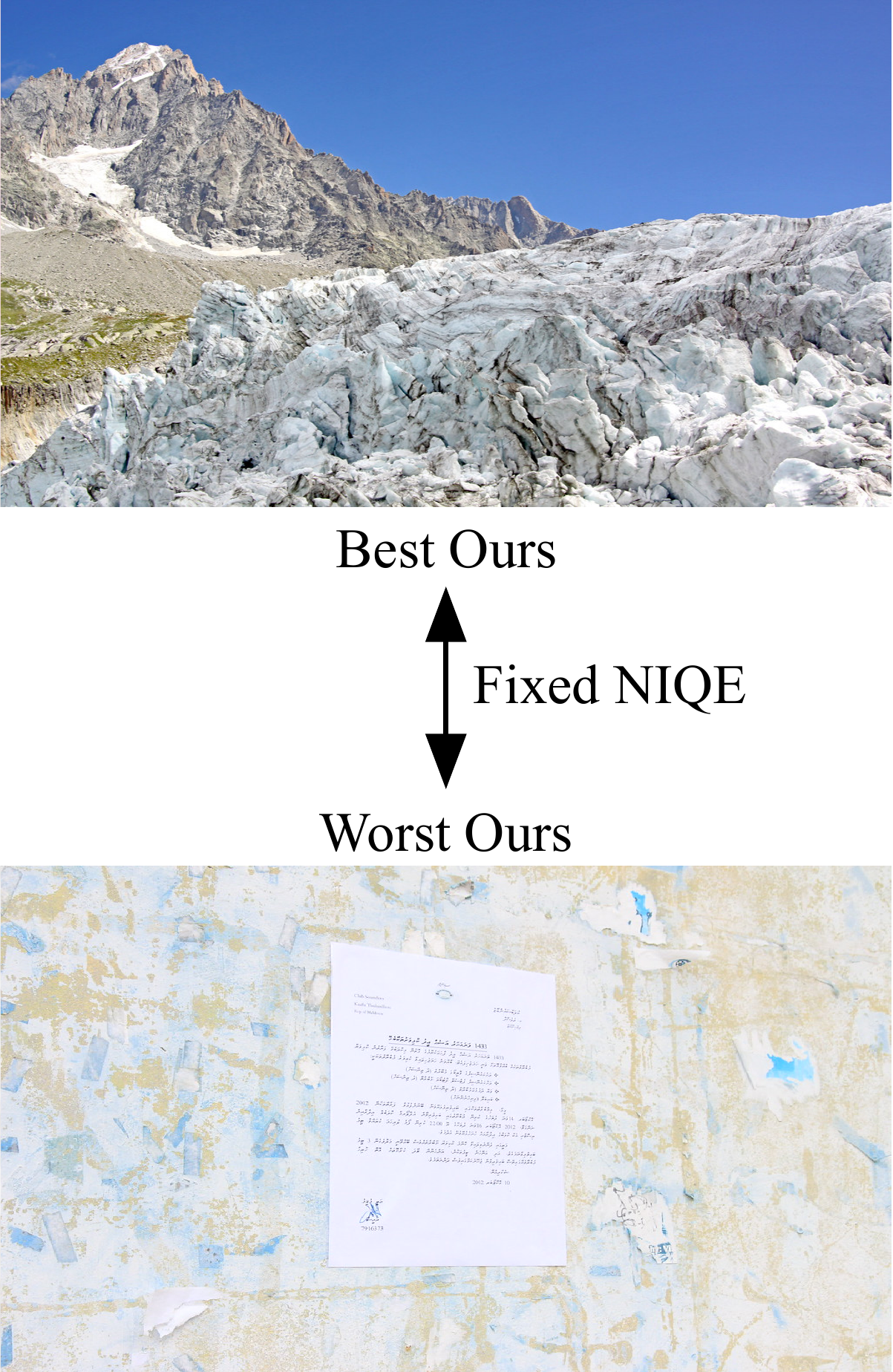}}\hskip.25em
	\subfloat[]{\includegraphics[width=0.23\textwidth]{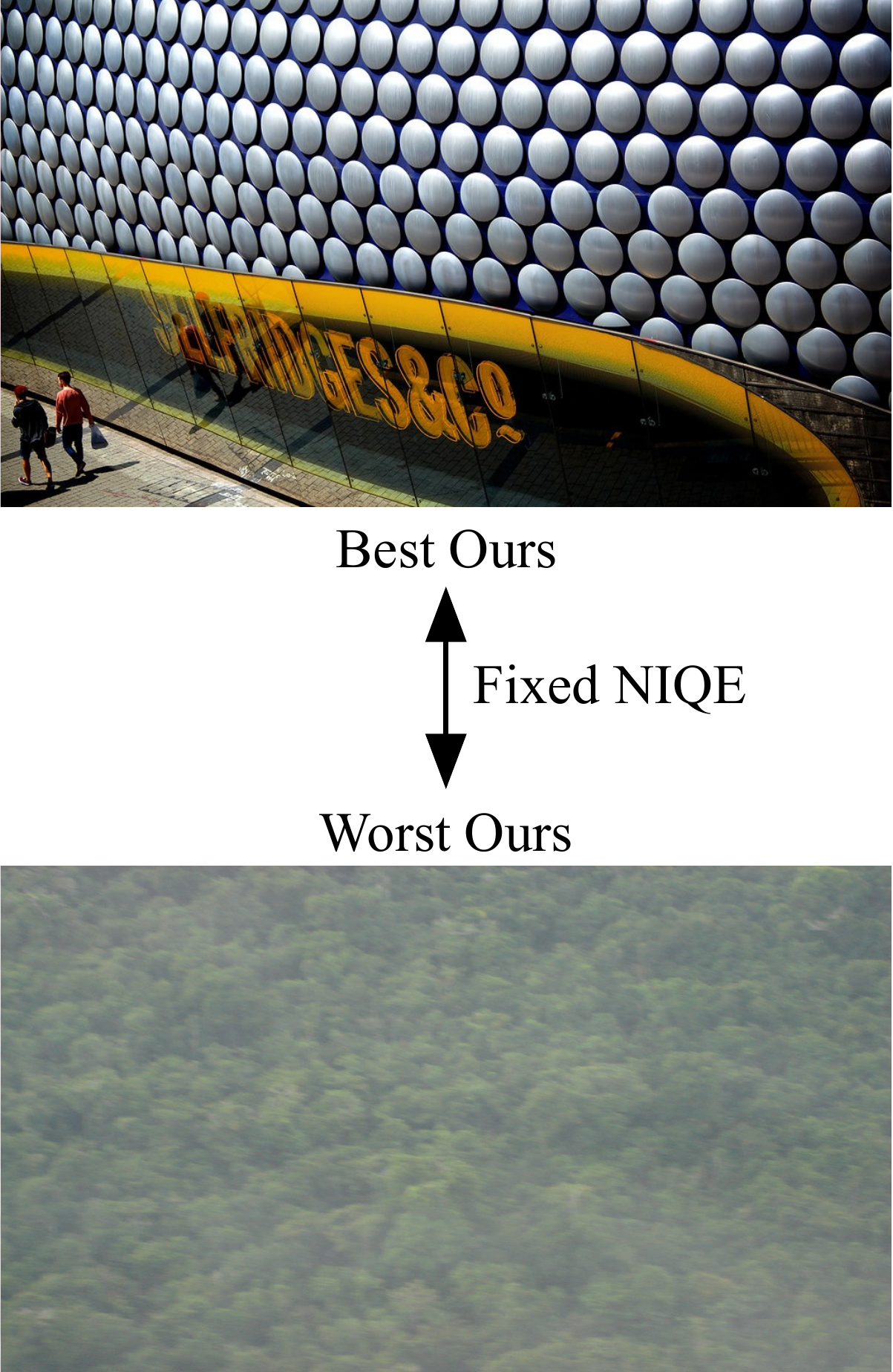}}\hskip.25em
	\subfloat[]{\includegraphics[width=0.23\textwidth]{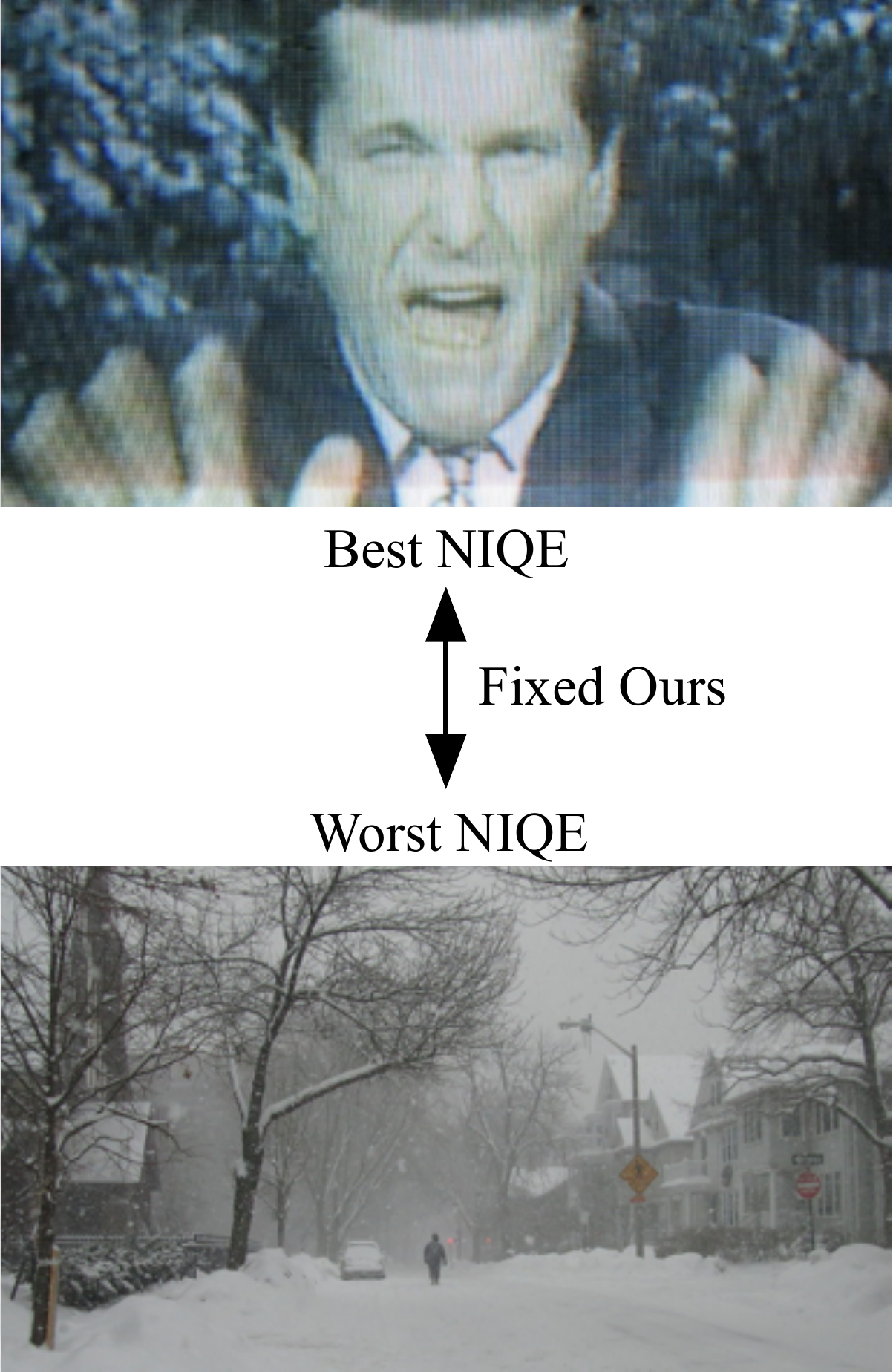}}\hskip.25em
	\subfloat[]{\includegraphics[width=0.23\textwidth]{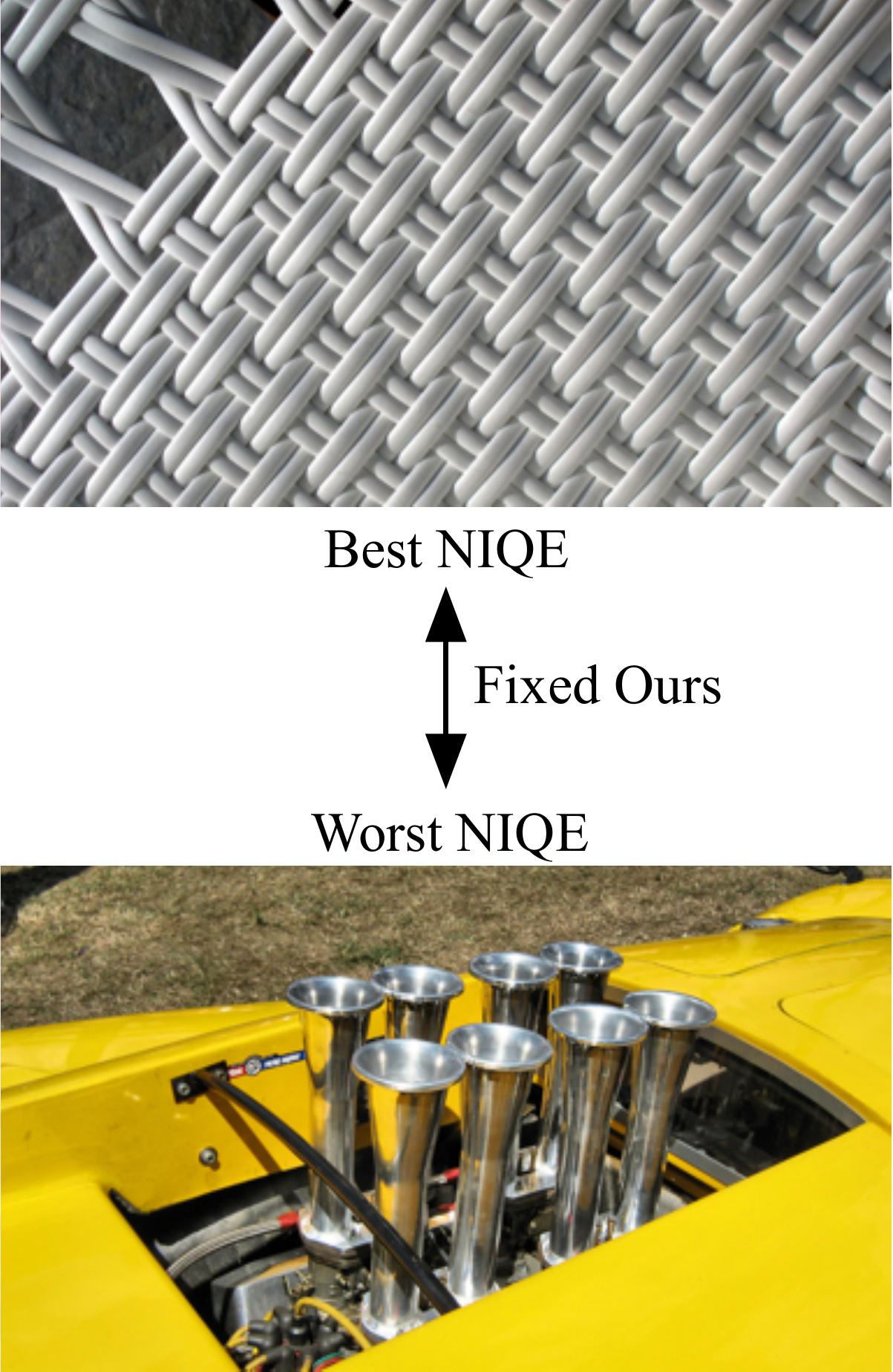}}
	\caption{Representative gMAD pairs between our method and NIQE when $\mathcal{X}_t$ is KonIQ-10k \cite{hosu2020koniq}.
	\textbf{(a)} Fixing NIQE at the low quality level. 
    % 	From top to bottom, the MOSs are $51$ and $18$, respectively. 
    \textbf{(b)} Fixing NIQE at the high quality level. 
    % From top to bottom, the MOSs are $76$ and $17$, respectively.
    \textbf{(c)} Fixing ours at the low quality level. 
    % From top to bottom, the MOSs are $32$ and $32$, respectively.
    \textbf{(d)} Fixing ours at the high quality level. % \textbf{(jz: fixed or fixing, not clear here?)}
    % From top to bottom, the MOSs are $62$ and $75$, respectively.
    }
	\label{fig:koniq_niqe}
\end{figure*}

\begin{figure*}[t]
	\centering
	% \captionsetup{justification=centering}
	\subfloat[]{\includegraphics[width=0.23\textwidth]{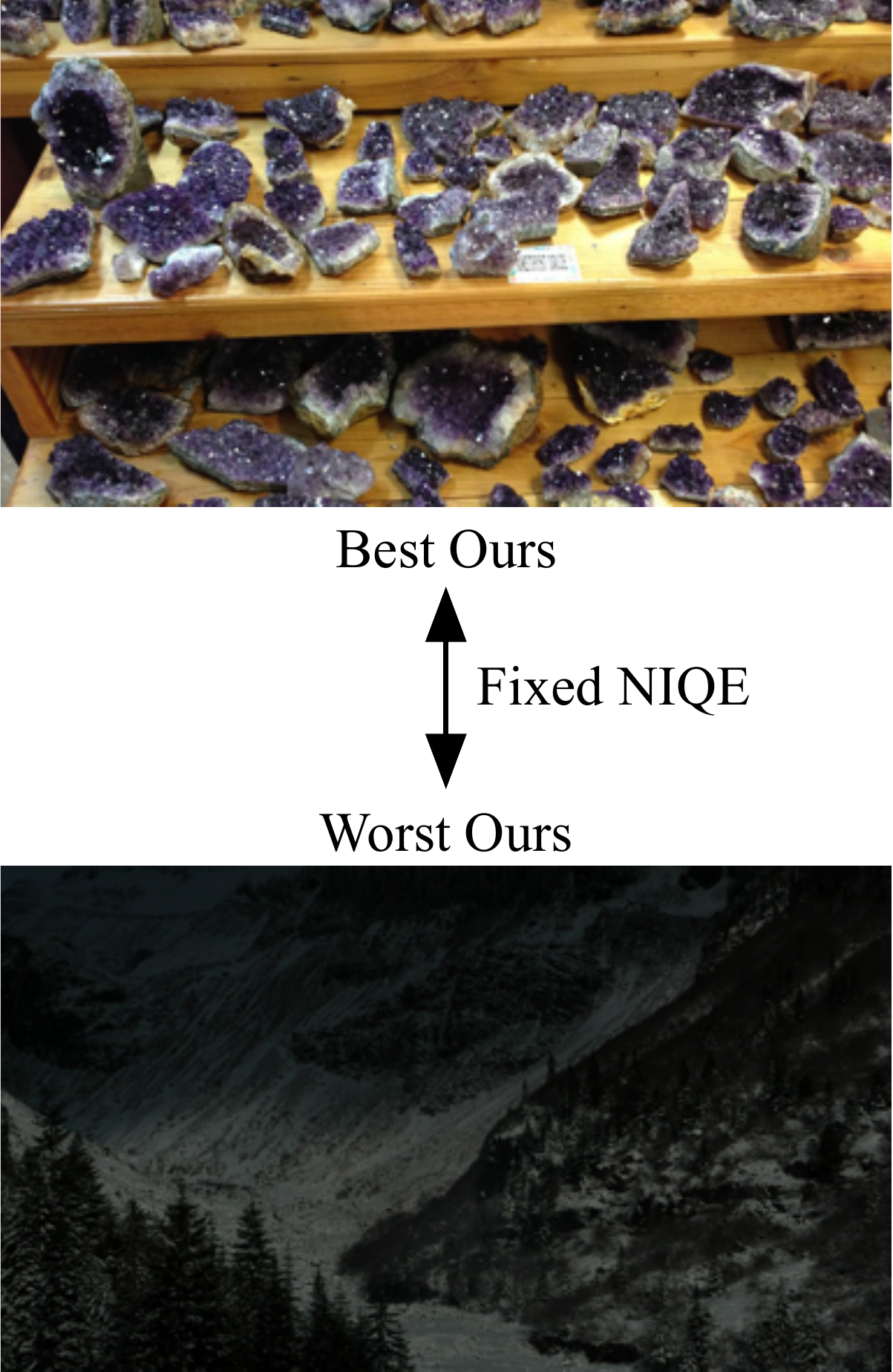}}\hskip.25em
	\subfloat[]{\includegraphics[width=0.23\textwidth]{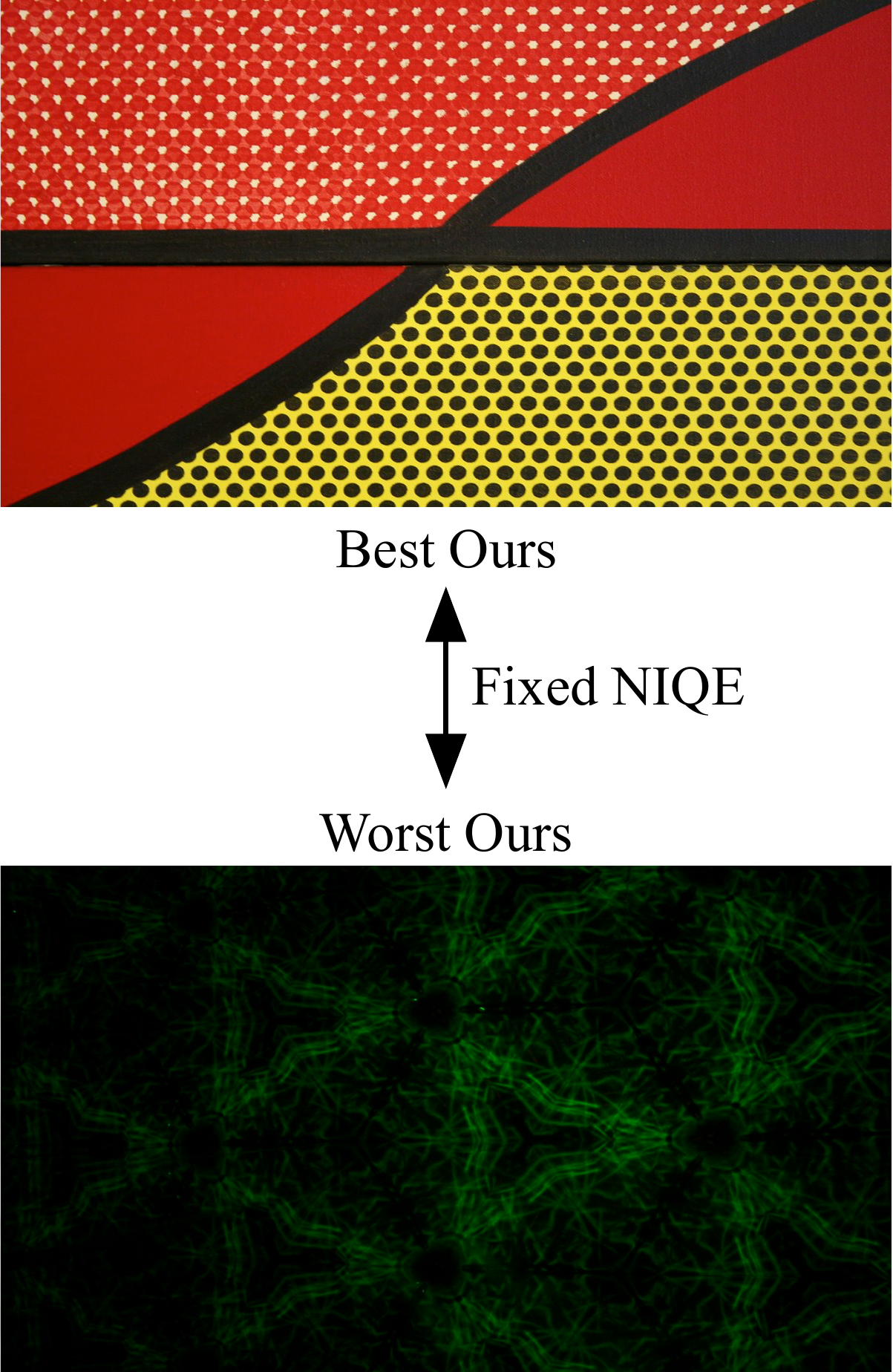}}\hskip.25em
	\subfloat[]{\includegraphics[width=0.23\textwidth]{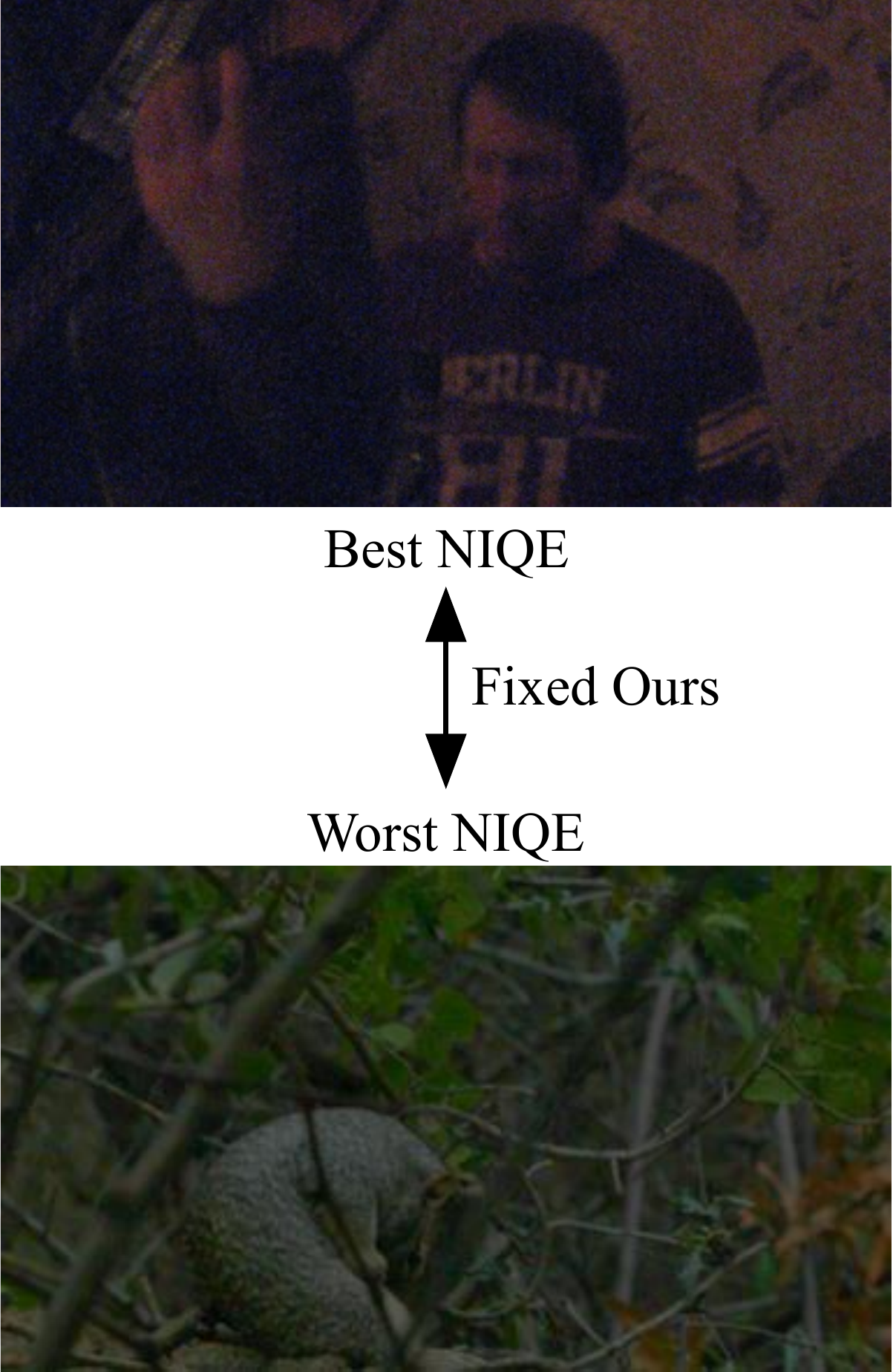}}\hskip.25em
	\subfloat[]{\includegraphics[width=0.23\textwidth]{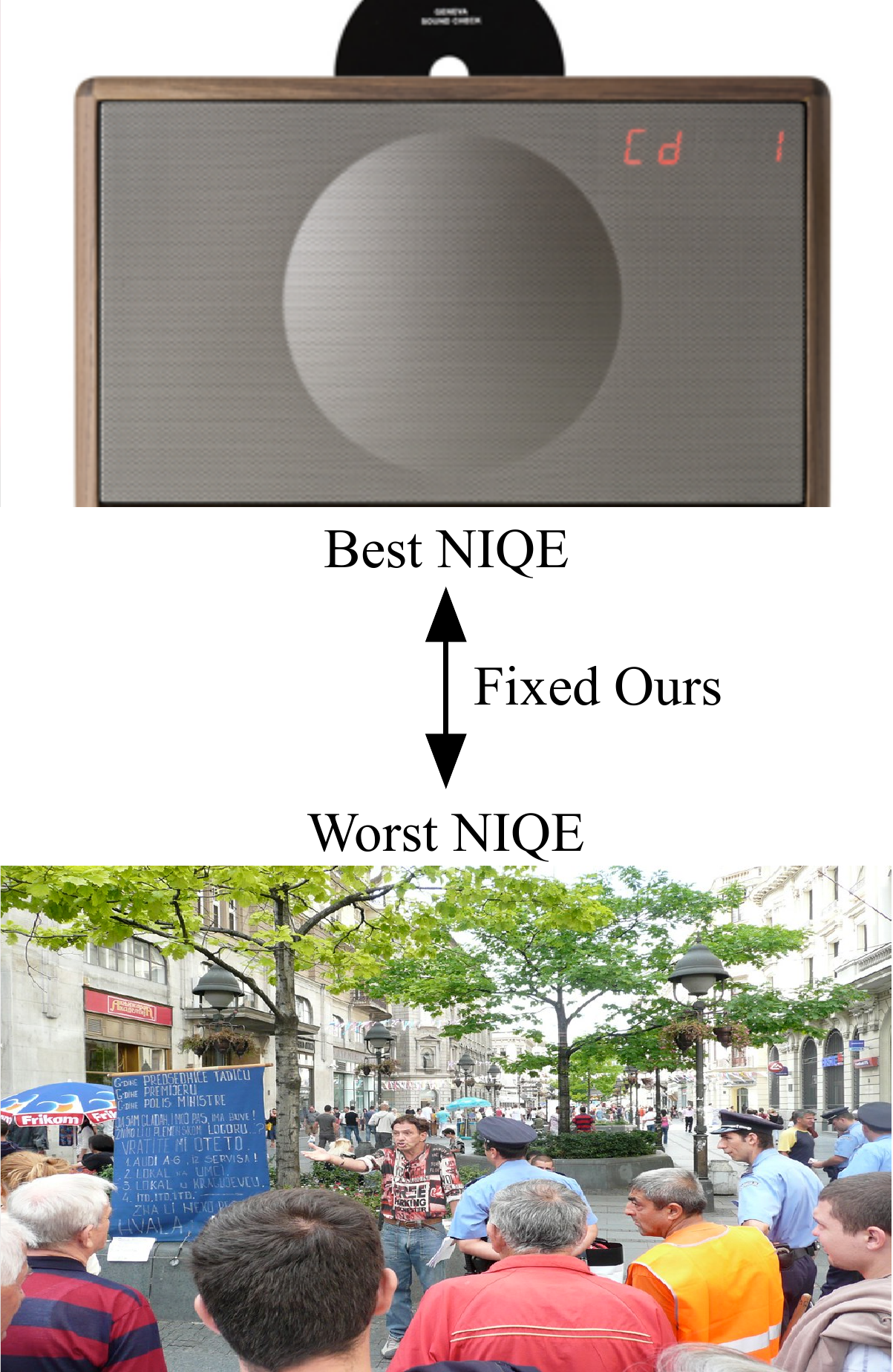}}
	\caption{Representative gMAD pairs between our method and NIQE when $\mathcal{X}_t$ is SPAQ \cite{fang2020cvpr}. 
	\textbf{(a)} Fixing NIQE at the low quality level. 
    % 	From top to bottom, the MOSs are $51$ and $18$, respectively. 
    \textbf{(b)}  Fixing NIQE at the high quality level. 
    % From top to bottom, the MOSs are $76$ and $17$, respectively.
    \textbf{(c)} Fixing ours at the low quality level. 
    % From top to bottom, the MOSs are $32$ and $32$, respectively.
    \textbf{(d)} Fixing ours at the high quality level. 
    % From top to bottom, the MOSs are $62$ and $75$, respectively.
}
	\label{fig:spaq_niqe}
\end{figure*}
\subsection{Construction of $\mathcal{D}$}
\label{sec:synthesizing}
\subsubsection{Synthesizing authentic distortions}
As mentioned before, we simulate the source training set $\mathcal{X}_s$ by taking each image from Waterloo Exploration Dataset \cite{ma2016waterloo} as a reference. As discussed by Ghadiyaram \textit{et al.} \cite{ghadiyaram2015massive}, the authentic distortions could be seen as a mixture of several degradations in an unknown order that are inherently caused during image acquisition, processing, transmission, storage, etc. (see Fig. \ref{fig:syn_aut}). Therefore, the way to simulating the authentic distortions is of vital importance. After careful observations, it was found that out-of-focus, motion blur, compression artifacts, sensor noise, overexposure, underexposure, vignetting, and color changes are the dominating distortions in several existing authentically-distorted IQA datasets, \textit{e.g.}, LIVE Challenge \cite{ghadiyaram2015massive}, KonIQ-10k \cite{kong2018no} and SPAQ \cite{fang2020cvpr}. Based on this fact, we include ten types of synthetic distortions with five levels of severity to simulate the authentic distortions, \textit{i.e.}, Gaussian blur (simulate out-of-focus), motion blur, JPEG compression (simulate compression artifacts), JP2K compression (simulate compression artifacts),  Gaussian noise (simulate sensor noise), overexposure, underexposure, vignetting, chromatic aberration (simulate color changes) and contrast decrements (simulate color changes). We deform each pristine image in four ways, $i.e.,$ with (1) a single distortion, (1) random combination of two distortions, (3) random combination of three distortions, and (4) random combination of four distortions, where the distortion order is random excluding (1). We yield a total of $4744 \times  50 = 273,200$ images, $i.e.,$ $n_s = 273,200$ (the proportions of four categories are $40\%$, $30\%$, $20\%$, and $10\%$, respectively)
%\textbf{(JZ: what is the meaning of 50 here?)}.
Fig.~\ref{fig:syn_aut} illustrates representative simulated images and authentically degraded images with similar distortion types. From those images, we could observe that the simulated distortions are perceptually similar to the corresponding authentic ones.

\subsubsection {Generating pseudo binary labels}
For a fair comparison, we randomly generate four types of image pairs ($x, y$), following the same protocol in \cite{ma2019blind}. Specifically, the two images in a pair present (1) different distortion levels of the same distortion types applied to the same reference image; (2) different distortion types and levels, but applied to the same reference; (3) different distortion types, levels and references; (4) one distorted and the other undistorted, both from the same reference image. After distortion, we generate about $426,684$ image pairs, $i.e.,$ $n = 426,684$ (the proportions of four categories are about $11\%$, $49\%$, $28\%$, and $12\%$, respectively). We use six state-of-the-art FR-IQA models as agents without training on human-rated IQA datasets to assign the pseudo-binary labels to each pairs in $\mathcal{D}$, $i.e.,$ FSIM$_c$ \cite{zhang2011fsim}, SR-SIM \cite{zhang2012sr}, NLPD \cite{laparra2017perceptually}, VSI \cite{zhang2014vsi}, MDSI \cite{nafchi2016mean} and GMSD \cite{xue2013gradient}, due to their excellent performance. Implementations of all six models were obtained from the respective authors' released codes with default parameters. To test the consensus of all six FR-IQA models, we report the annotation consistency on $\mathcal{D}$ in Fig. \ref{fig:distribution}. 
%\textbf{(JZ: this figure appears suddenly without any context. Where are the results from?)} 
We can easily conclude that approximately $80$\% of image pairs are in agreement with all models, indicating the rationality and reliability of these FR-IQA models in the quality assessment of synthetically-distorted images.

\begin{figure*}[t]
	\centering
	% \captionsetup{justification=centering}
	\subfloat[]{\includegraphics[width=0.23\textwidth]{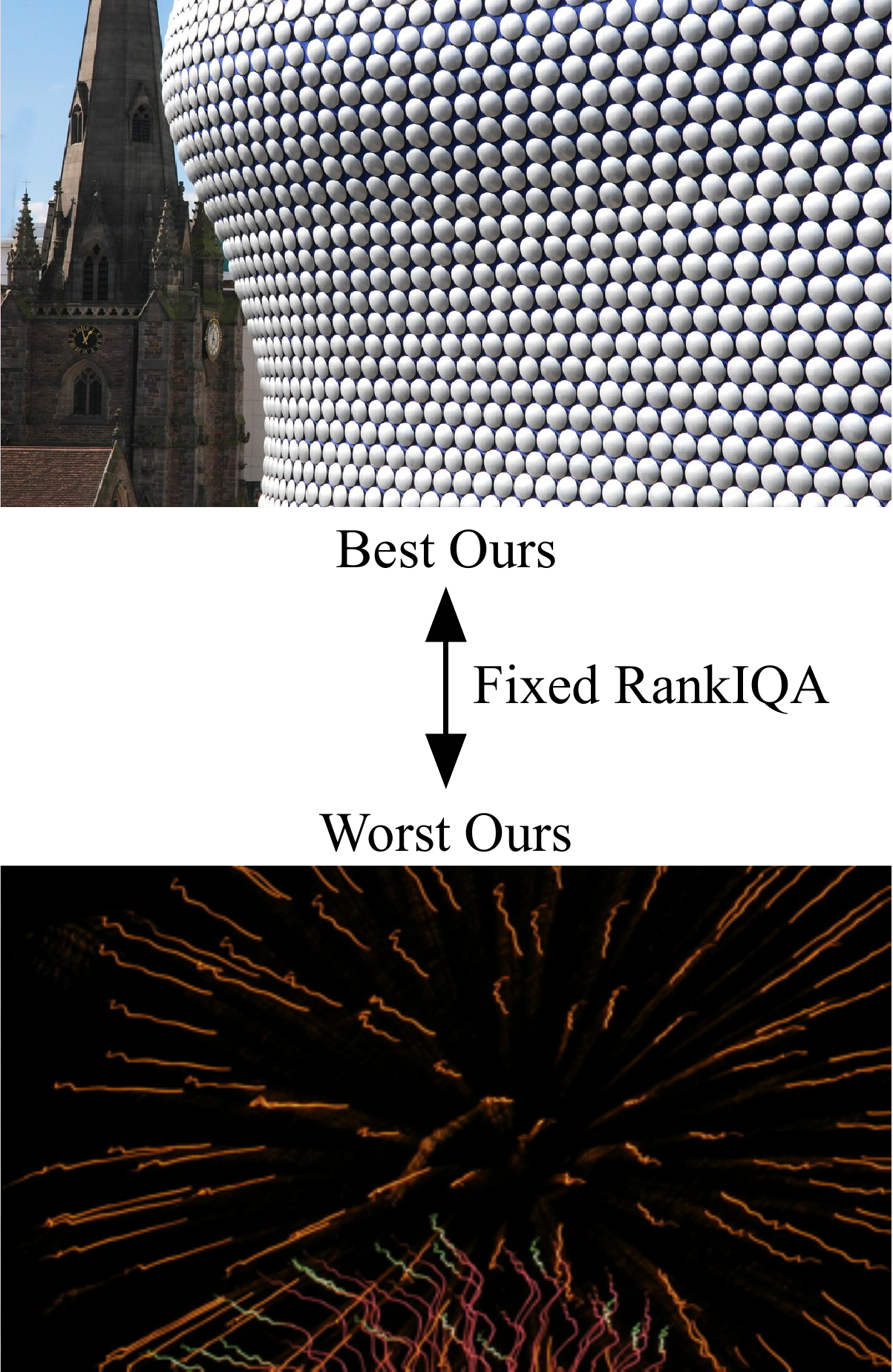}}\hskip.25em
	\subfloat[]{\includegraphics[width=0.23\textwidth]{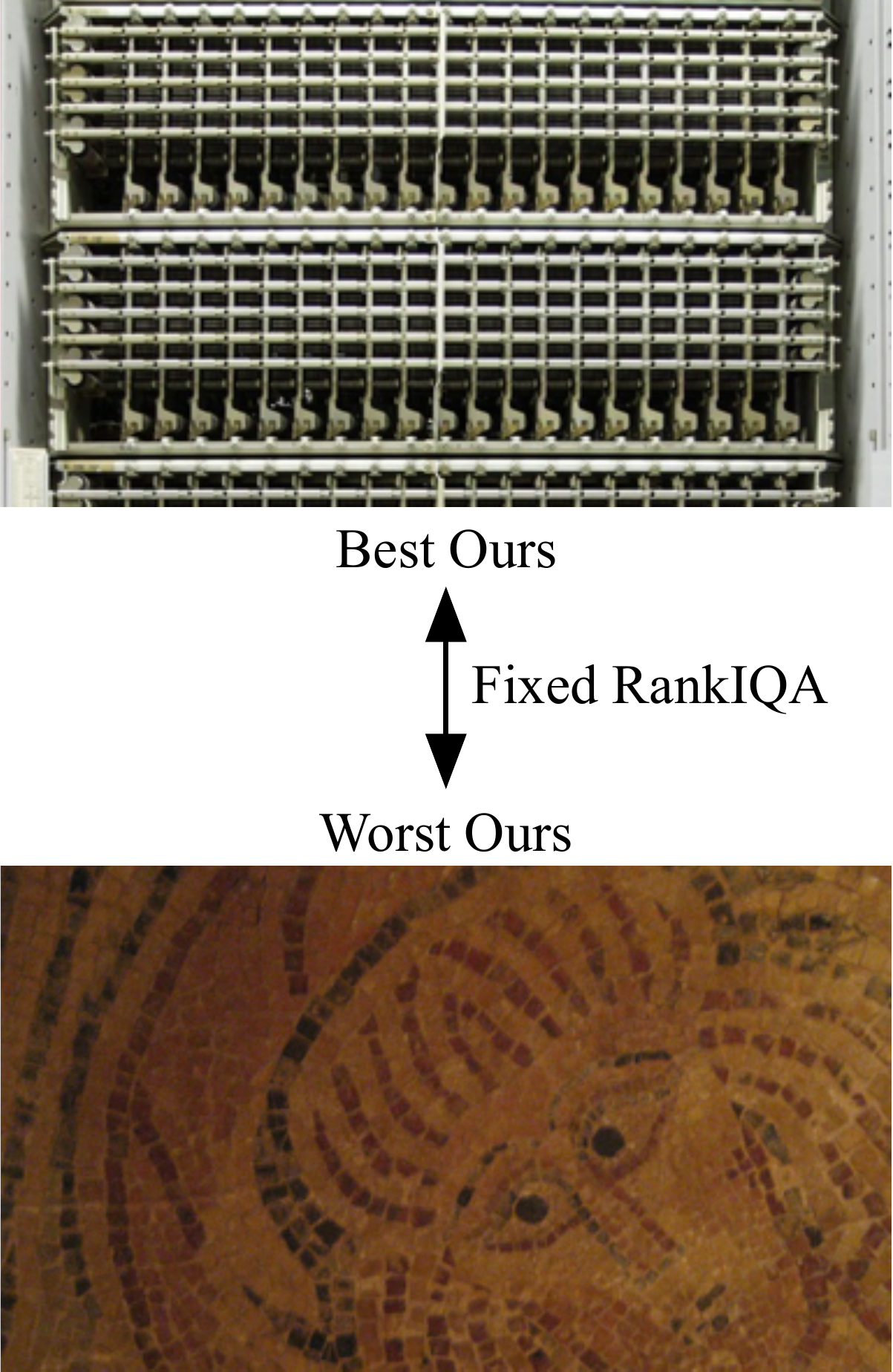}}\hskip.25em
	\subfloat[]{\includegraphics[width=0.23\textwidth]{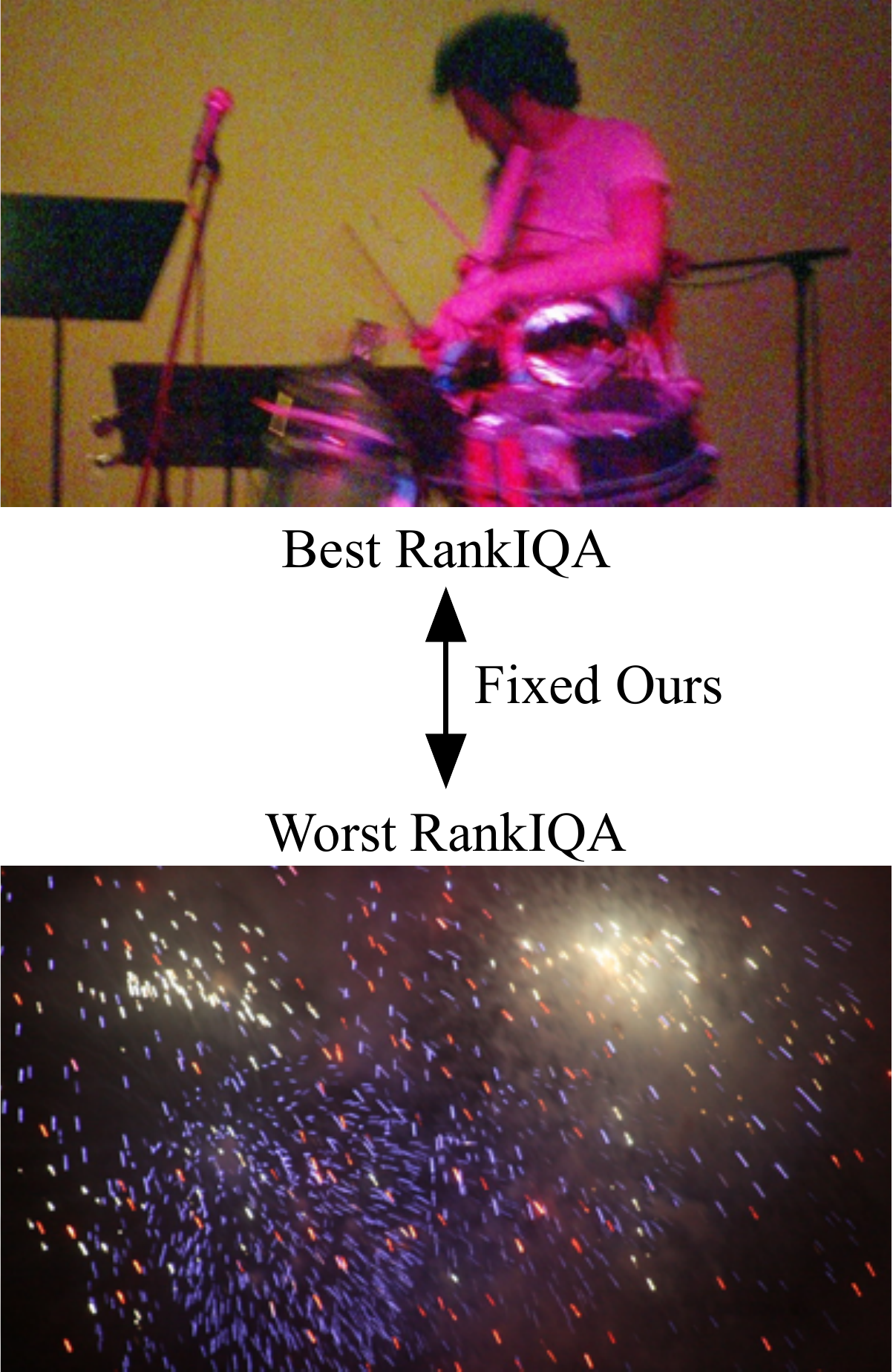}}\hskip.25em
	\subfloat[]{\includegraphics[width=0.23\textwidth]{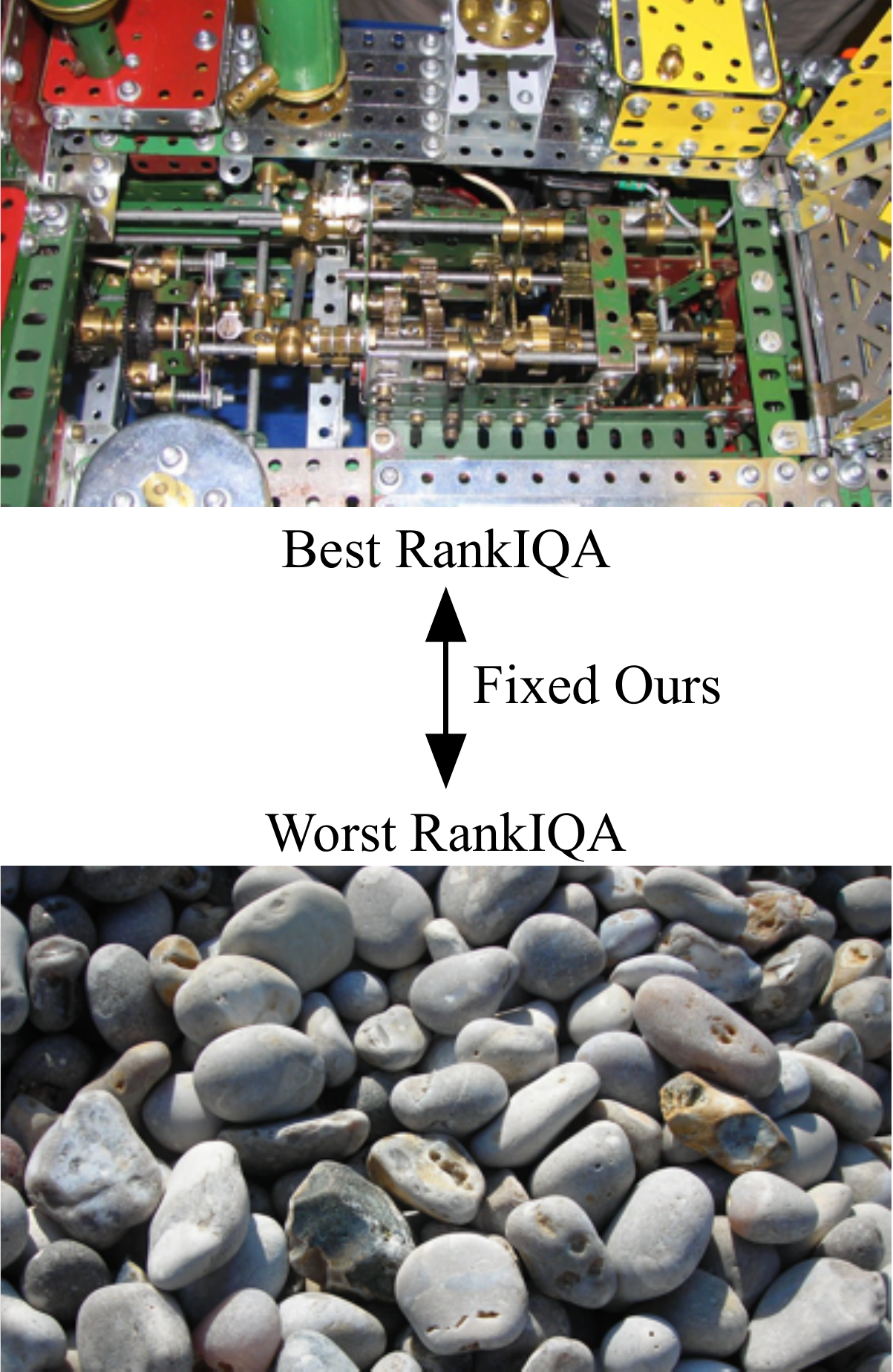}}
	\caption{Representative gMAD pairs between ours and RankIQA \cite{liu2017rankiqa} when $\mathcal{X}_t$ is KonIQ-10k \cite{hosu2020koniq}. 
	\textbf{(a)} Fixing RankIQA at the low quality level. 
    % 	From top to bottom, the MOSs are $51$ and $18$, respectively. 
    \textbf{(b)}  Fixing RankIQA at the high quality level. 
    % From top to bottom, the MOSs are $76$ and $17$, respectively.
    \textbf{(c)} Fixing ours at the low quality level. 
    % From top to bottom, the MOSs are $32$ and $32$, respectively.
    \textbf{(d)} Fixing ours at the high quality level. 
    % From top to bottom, the MOSs are $62$ and $75$, respectively.
    }
	\label{fig:koniq_rankiqa}
\end{figure*}

\begin{figure*}[t]
	\centering
	% \captionsetup{justification=centering}
	\subfloat[]{\includegraphics[width=0.23\textwidth]{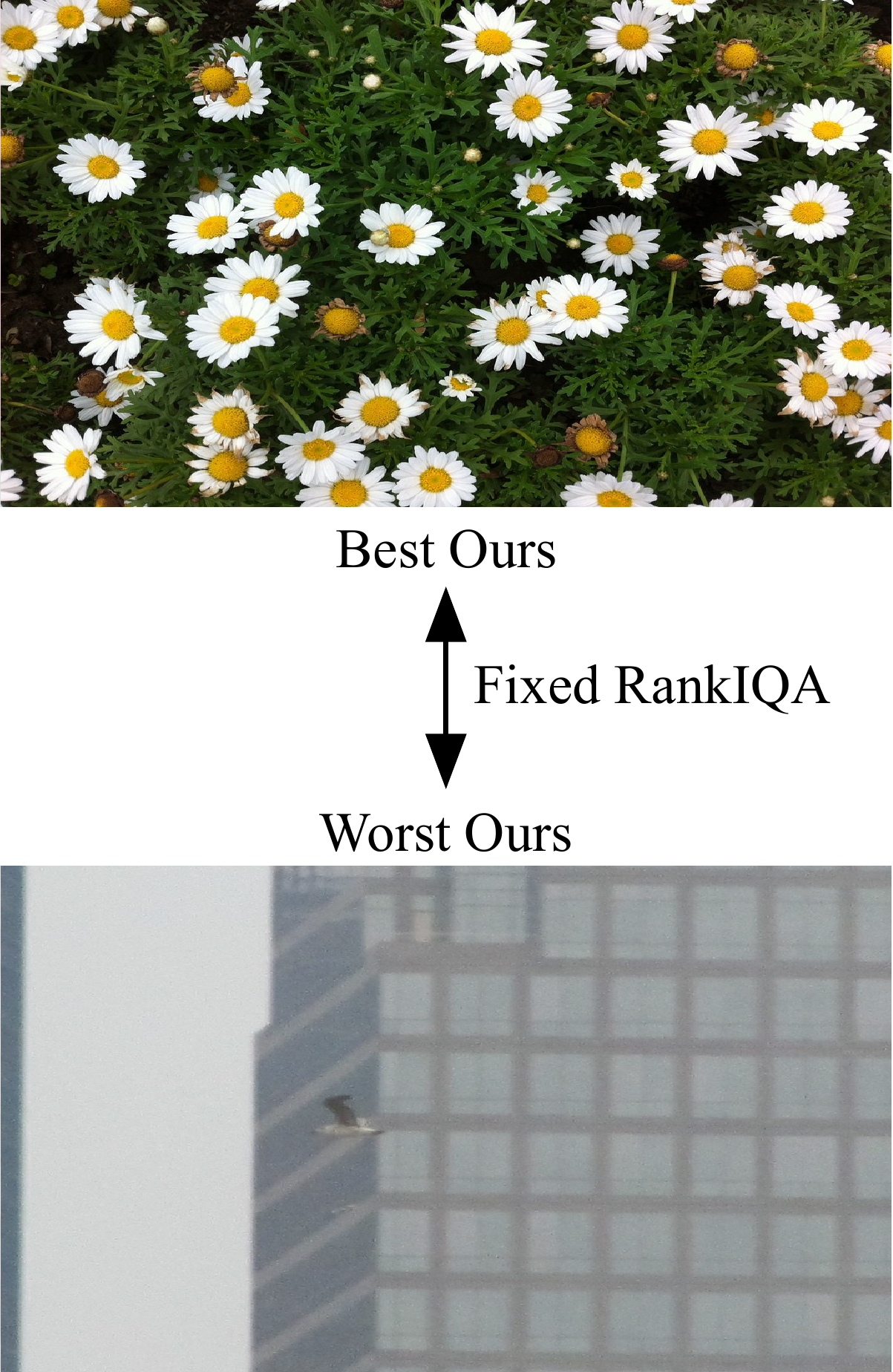}}\hskip.25em
	\subfloat[]{\includegraphics[width=0.23\textwidth]{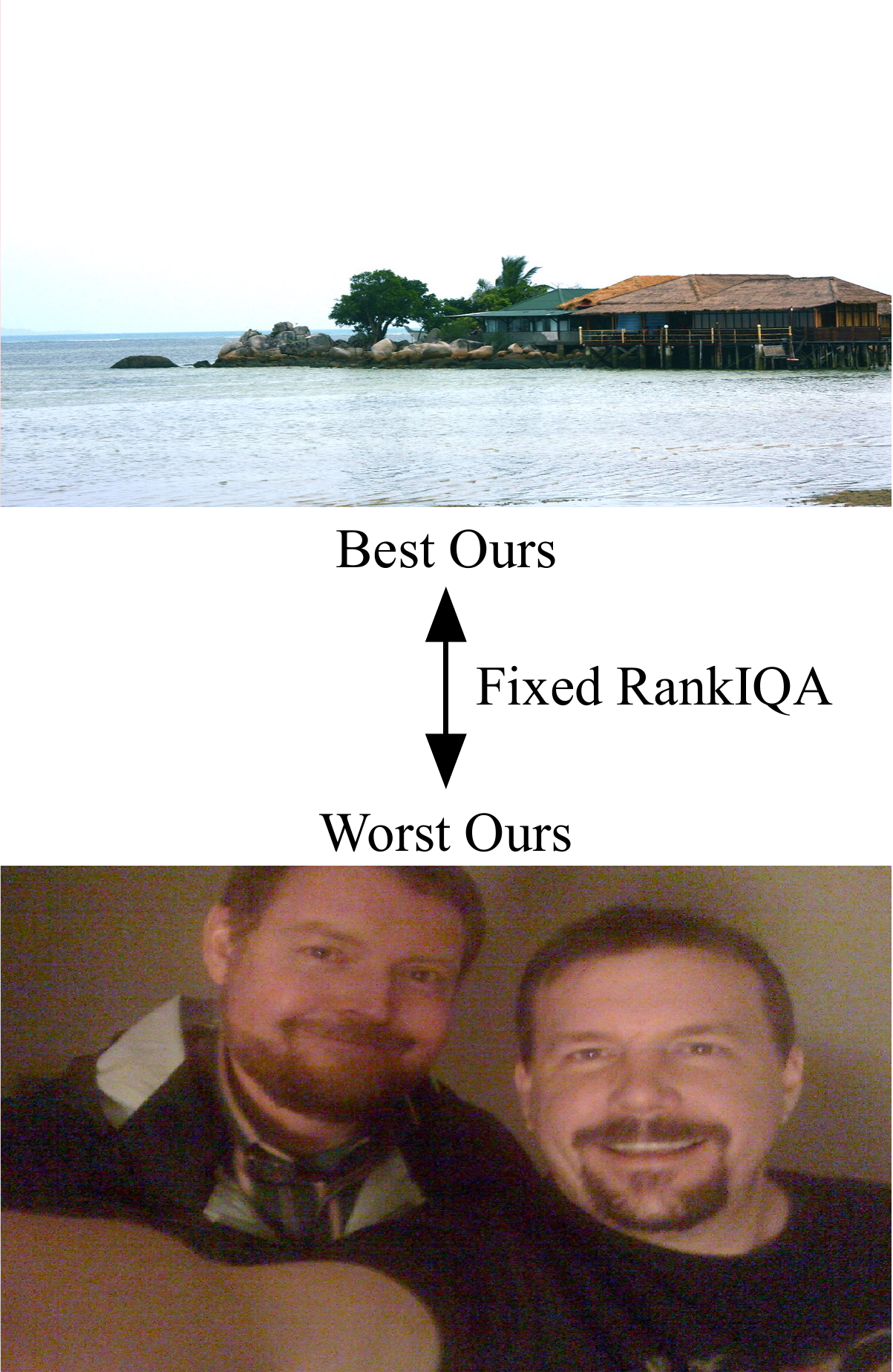}}\hskip.25em
	\subfloat[]{\includegraphics[width=0.23\textwidth]{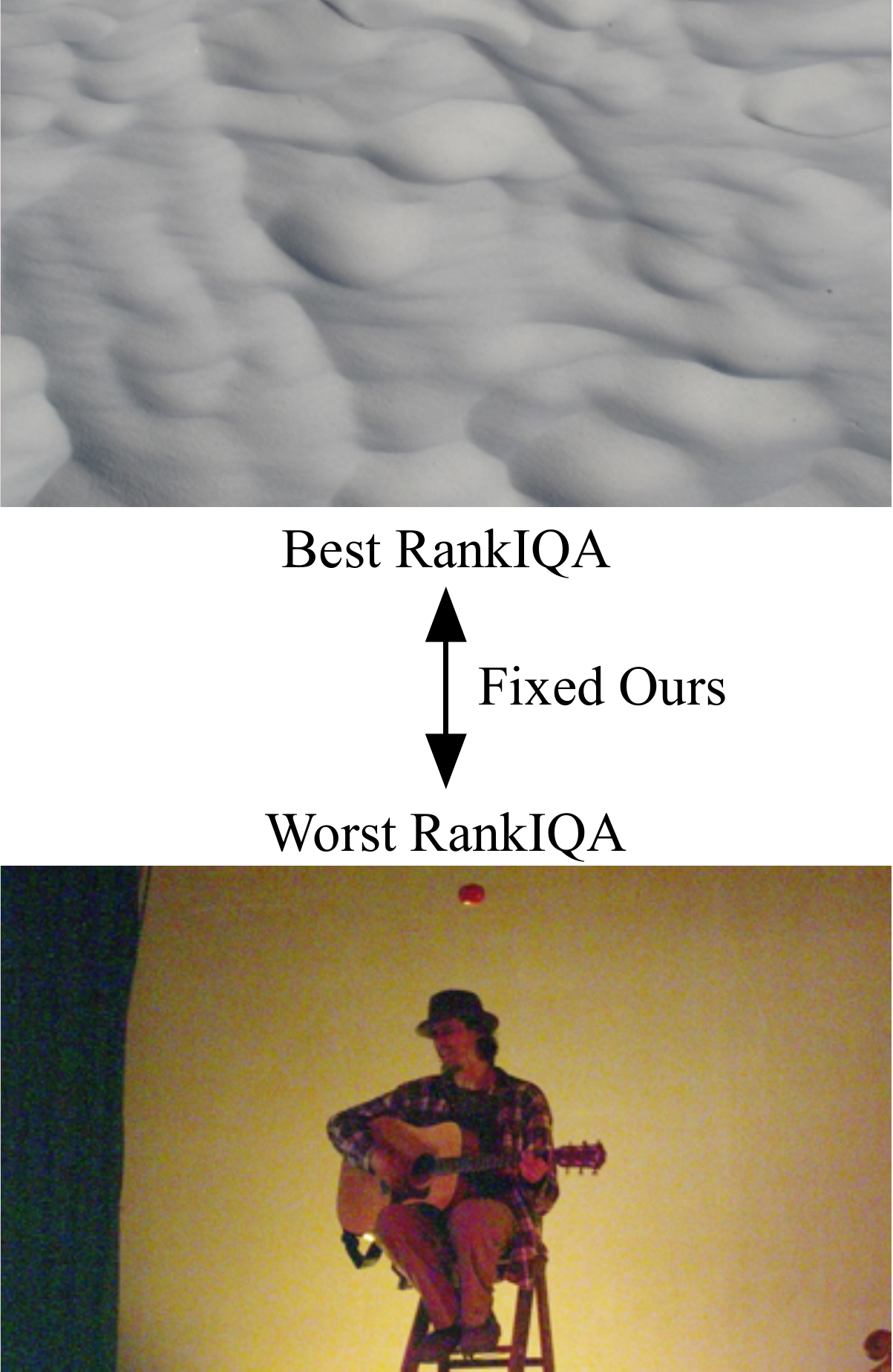}}\hskip.25em
	\subfloat[]{\includegraphics[width=0.23\textwidth]{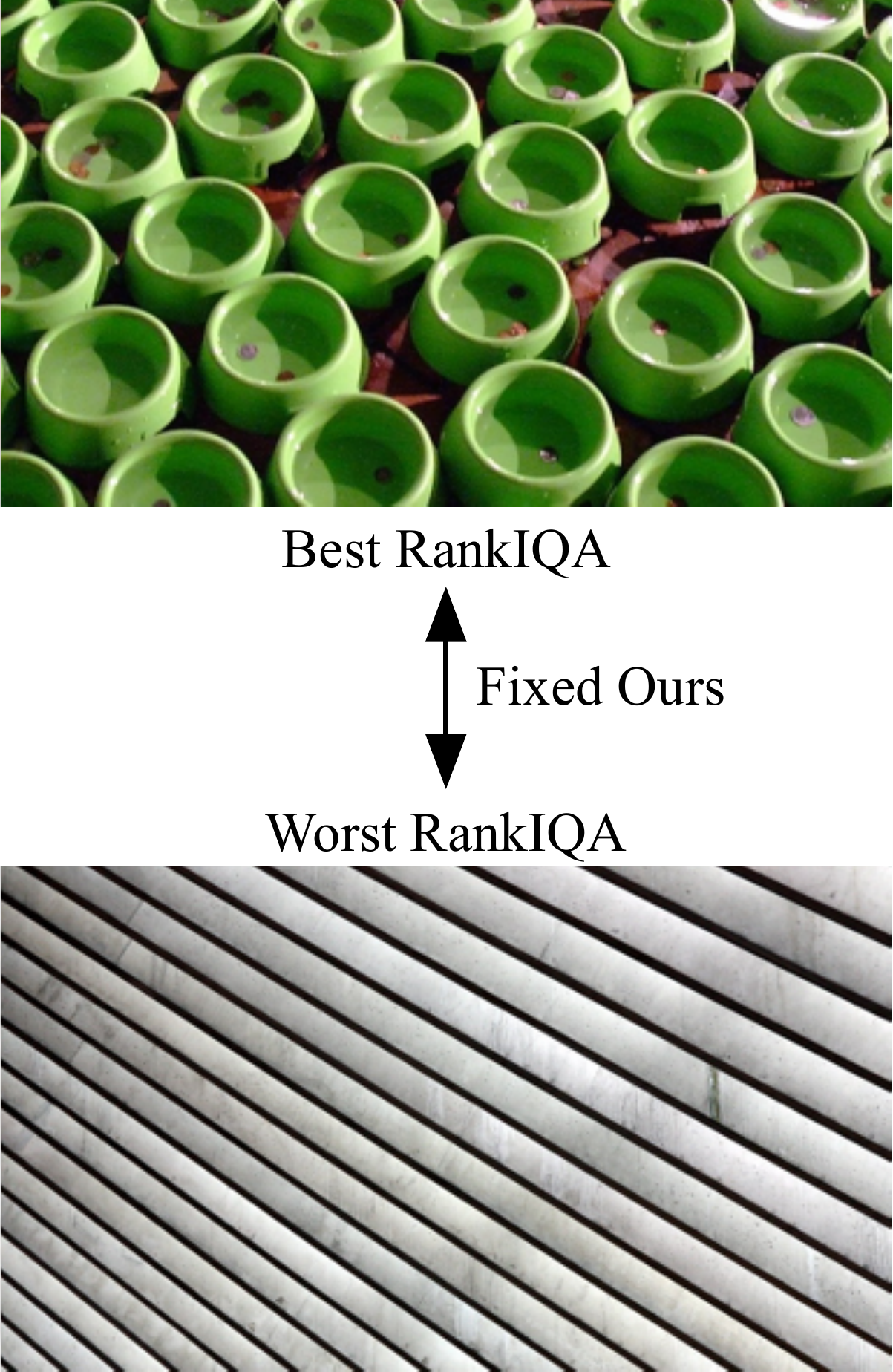}}
	\caption{Representative gMAD pairs between ours and RankIQA \cite{liu2017rankiqa} when $\mathcal{X}_t$ is SPAQ \cite{fang2020cvpr}. 
	\textbf{(a)} Fixing RankIQA at the low quality level. 
    % 	From top to bottom, the MOSs are $51$ and $18$, respectively. 
    \textbf{(b)}  Fixing RankIQA at the high quality level. 
    % From top to bottom, the MOSs are $76$ and $17$, respectively.
    \textbf{(c)} Fixing ours at the low quality level. 
    % From top to bottom, the MOSs are $32$ and $32$, respectively.
    \textbf{(d)} Fixing ours at the high quality level. 
    % From top to bottom, the MOSs are $62$ and $75$, respectively.
    }
	\label{fig:spaq_rankiqa}
\end{figure*}

\subsection {Experimental Details}
The specification of feature extractor $F$, quality predictor $Q$, domain discriminator $D$, and the consensus learning classifier $C$ is illustrated in Fig. \ref{fig:network}. The backbone of the feature extractor $F$ used in our experiment is ResNet-18 \cite{he2016identity}. % due to its good balance between model complexity and representation ability
The quality predictor $Q$ consists of three fully-connected layers and takes Leaky RELU as the activation function. The outputs of $Q$ are two predicted values - image quality $f_{w_q}(x)$ and corresponding STD $\sigma_{w_q}(x)$ for an input $x$. To constrain the STD $\sigma_{w_q}(x)$ to be positive, a softplus function is applied on $\sigma_{w_q}(x)$. Similarly, the domain discriminator and CLC are also composed using three fully-connected layers while taking RELU nonlinearity as activation function with one output representing domain label and the other for the ranking consensus, respectively. 

\begin{table*}
\caption{Summary of the learned values of hit rate $\alpha$ and correct rejection rate $\beta$ of each FR-IQA agents.}
\centering
\begin{threeparttable}
\begin{tabular}{l|cccccc}
\toprule
$\alpha$ & FSIM$_c$ \cite{zhang2011fsim} & SR-SIM \cite{zhang2012sr} & NLPD \cite{laparra2017perceptually} & VSI \cite{zhang2014vsi} &  MDSI \cite{nafchi2016mean} & GMSD \cite{xue2013gradient}\\
\hline
Ours(na\"ive) & $0.697$ & $0.700$ & $0.699$ & $0.581$ & $0.703$ & $0.659$\\
Ours(KonIQ-10k \cite{hosu2020koniq}) & $0.700$ & $0.703$ & $0.703$ & $0.585$ & $0.706$ & $0.663$\\
Ours(SPAQ \cite{fang2020cvpr}) & $0.700$ & $0.703$ & $0.703$ & $0.585$ & $0.706$ & $0.663$\\
\hline
\hline
$\beta$ & FSIM$_c$ & SR-SIM & NLPD & VSI &  MDSI & GMSD\\
\hline
Ours(na\"ive) & $0.640$ & $0.676$ & $0.681$ & $0.686$ & $0.720$ & $0.759$\\
Ours(KonIQ-10k) & $0.643$ & $0.681$ & $0.685$ & $0.689$ & $0.723$ & $0.762$\\
Ours(SPAQ) & $0.643$ & $0.681$ & $0.685$ & $0.689$ & $0.723$ & $0.762$\\
\bottomrule
\end{tabular}
\end{threeparttable}
\label{tab:alpha_beta}
\end{table*}

During training, $\lambda_1$, $\lambda_2$ and $\lambda_3$ were set to $0.08$, $0.08$ and $0.02$ respectively, which are consistent across all the experiments. We firstly pre-trained $F$ and $Q$ for $T_1$ = $8$ epochs, and then performed domain adaptation $T_2$ = $2$ extra epochs. The initial learning rate was set to $10^{-4}$ with a decay rate of $3$ for every three epochs, while the learning rate of domain adaptation was set to $10^{-6}$ remaining unchanged. Images were re-scaled and cropped into the size: $384\times384\times3$  and served as inputs during training, while the model was tested on images of full size. Codes were written using PyTorch, and the experiments were conducted using a single NVIDIA Titan V100 GPU.% \textbf{(JZ: please comment\ref on the time taken to train the model, i.e,, the computational cost.)}  

\subsection{Evaluation Criteria}
Two evaluation metrics are used to quantitatively evaluate the performance of IQA methods, $i.e.$, the Spearman's rank correlation coefficient (SRCC) and the Pearson linear correlation coefficient (PLCC). SRCC measures the prediction monotonicity, while PLCC measures the prediction precision between MOSs and the predicted quality scores. As suggested in \cite{video2000final}, a pre-processing step is added to linearize model predictions by fitting a four-parameter monotonic function before computing PLCC
\begin{align}
    \hat{f}_{w_q}(x) = (\eta_1 -\eta_2)/(1+\exp(-(f_{w_q}(x)-\eta_3)/|\eta_4|)) + \eta_2.
    \label{eq:nm}
\end{align} 
where $\{\eta_i; i=1,2,3,4\}$ are the regression parameters to be fitted.

To evaluate the generalizability and robustness of our proposed method, we also utilized group MAximum Differentiation (gMAD) competition \cite{ma2016group, ma2018group, zhang2018blind, zhang2020uncertainty}, a widely adopted strategy for benchmarking IQA models on large-scale unlabeled datasets without human ratings. Intuitively, gMAD competition focuses on the pair of images that can falsify perceptual models in the most efficient way. Given two models, gMAD seeks a pair of images with similar quality scores predicted by one model while being assessed substantially differently by the competing model. By comparing the selected image pairs, we can easily observe the differences between two models in the error-spotting capacity, qualitatively indicating the generalizability and robustness when they work in a real-world application.

\subsection {Results}

\renewcommand{\multirowsetup}{\centering}
\begin{table}[t]
    \caption{Effect of distortion mixture: results of simulating $\mathcal{X}_s$ with single distortions.}
    \label{table:ablation_single_mixture}
    \vspace{-.3cm}
	\begin{center}
		\begin{tabular}{l|ccc}
    		\toprule[1pt]
			SRCC & LIVE\cite{sheikh2006statistical} & KonIQ-10k\cite{hosu2020koniq} & SPAQ\cite{fang2020cvpr} \\
			\hline
			%$-$ & $0.914$ & $0.618$ & $0.715$\\
			Ours(KonIQ-10k) & $0.905$ & $0.640$ & $0.753$\\
			Ours(SPAQ) & $0.912$ & $0.631$ & $0.749$\\
			\hline
			\hline
			PLCC  & LIVE & KonIQ-10k & SPAQ\\
			\hline
			%$-$ & $0.910$ & $0.649$ & $0.724$\\
			Ours(KonIQ-10k) & $0.900$ & $0.676$ & $0.760$ \\
			Ours(SPAQ) & $0.910$ & $0.672$ & $0.756$ \\
			\bottomrule[1pt]
		\end{tabular}
	\end{center}
\end{table}

\renewcommand{\multirowsetup}{\centering}
\begin{table*}[t]
    \caption{Effects of different components and their combinations in our proposed method, including BCE, ADL, CLC and Mixup. For each metric, the best two results are highlighted in boldface}
    \label{table:ablation}
    \vspace{-.3cm}
	\begin{center}
		\begin{tabular}{l|cccc|ccc}
    		\toprule[1pt]
			SRCC  & BCE & ADL & CLC & Mixup & LIVE \cite{sheikh2006statistical} & KonIQ-10k \cite{hosu2020koniq} & SPAQ
			\cite{fang2020cvpr} \\
			\hline
			\multirow{3}{*}[-6pt]{Ours(KonIQ-10k)} & $\checkmark$ & & & &$\textbf{0.911}$ & $0.659$ & $0.825$ \\
			& & $\checkmark$ & & & $0.906$ & $0.670$ & $0.825$\\
			& & $\checkmark$ & $\checkmark$ & & $0.907$ & $0.707$ & $0.827$\\
			& & $\checkmark$ & $\checkmark$ & $\checkmark$ & $0.907$ & $\textbf{0.717}$ & $0.826$ \\
			\hline
			\multirow{3}{*}[-6pt]{Ours(SPAQ)} & $\checkmark$ & & & & $0.885$ & $0.685$ & $0.822$\\
			& & $\checkmark$ & & & $0.880$ & $0.689$ & $0.828$\\
			& & $\checkmark$ & $\checkmark$ & & $0.906$ & $0.702$ & $\textbf{0.834}$\\
			& & $\checkmark$ & $\checkmark$ & $\checkmark$ & $\textbf{0.920}$ & $\textbf{0.712}$ & $\textbf{0.838}$\\
			\hline
			\hline
			PLCC  & BCE & ADL & CLC & Mixup & LIVE & KonIQ-10k & SPAQ\\
			\hline
			\multirow{3}{*}[-6pt]{Ours(KonIQ-10k)} & $\checkmark$ & & & & $0.904$ & $0.672$ & $0.830$\\
			& & $\checkmark$ & & & $0.900$ & $0.684$ & $0.831$\\
			& & $\checkmark$ & $\checkmark$ & & $0.903$ & $0.726$ & $0.836$\\
			& & $\checkmark$ & $\checkmark$ & $\checkmark$ & $\textbf{0.906}$ & $\textbf{0.740}$ & $0.831$\\
			\hline
			\multirow{3}{*}[-6pt]{Ours(SPAQ)} & $\checkmark$ & & & & $0.872$ & $0.707$ & $0.825$\\
			& & $\checkmark$ & & & $0.869$ & $0.709$ & $0.832$\\
			& & $\checkmark$ & $\checkmark$ & & $0.902$ & $0.729$ & $\textbf{0.840}$\\
			& & $\checkmark$ & $\checkmark$ & $\checkmark$ & $\textbf{0.912}$ & $\textbf{0.736}$ & $\textbf{0.844}$\\
			\bottomrule[1pt]
		\end{tabular}
	\end{center}
\end{table*}

\subsubsection{Quantitative comparisons} Our proposed method was compared to six state-of-the-art $opinion$-$free$ BIQA methods - QAC \cite{xue2013learning}, NIQE \cite{mittal2013making}, ILNIQE \cite{zhang2015feature}, RankIQA \cite{liu2017rankiqa}, dipIQ \cite{ma2017dipiq} and Ma19 \cite{ma2019blind}, where the first three models are knowledge-driven methods while the other three are data-driven methods. For those competing models, we used the publicly available default implementations provided by the corresponding authors for a fair comparison. TABLE \ref{table:comparision} summarizes the comparison results in terms of correlation (SRCC and PLCC), where Ours($\mathcal{X}_t$) with different $\mathcal{X}_t$ indicates our methods being adapted to different target domain, $e.g$. \textit{{Ours (SPQA)}} refers to our method with SPQA being used as target domain and \textit{Ours(na\"{i}ve)} is for our method with no target domain (those namings are made consistent across all of the experiments). We also include the performance on a synthetic dataset - LIVE \cite{sheikh2006statistical} as a reference to demonstrate that our models are able to handle both synthetic distortions and authentic distortions properly.  

There are several observations worth noting from the results in TABLE \ref{table:comparision}. First, both knowledge-driven and data-driven BIQA methods initially designed for synthetic distortions generally work poorly on authentic distortions; in particular, the QAC method catastrophically performs on the SPAQ and KonIQ-10k. It is as expected as there exists a significant distributional discrepancy between the two data distributions, and those methods considered didn't specifically seek to tackle this problem. Second, 
%NIQE and its improved version ILNIQE perform similarly well on realistic distortions compared with other competing models, suggesting the effectiveness of the handcrafted features at capturing the characteristics of authentic distortions. 
our proposed method outperforms six competing methods on authentic distortions even if no UDA. After UDA, a more obvious difference is observed. This indicates the effectiveness of our proposed method to learn a $opinion$-$free$ DNN-based BIQA model. Third, the performance on SPAQ is superior to that on KonIQ-10k. This phenomenon may be due to two main reasons: 1) SPAQ includes more easily-to-handle samples; 2) the distribution of $\mathcal{X}_s$ is close to SPAQ. Through the analysis of other competing models' performance on two datasets, we find that the first reason may be the dominant factor. Finally, whether SPAQ or KonIQ-10k works as $\mathcal{X}_t$, we can obtain a similar performance improvement on authentic distortions. This may be due to shared data distribution between two datasets. %\textbf{(JZ:the logic here is not clear, and please rewrite.)} 

As shown in Eq.(\ref{eq:alpha}) and Eq.(\ref{eq:beta}), the hit rate $\alpha$ and correct rejection rate $\beta$ are the learnable parameters indicating how much the learned network trusts those agents. To exam such confidences, we show the values of $\alpha$ and $\beta$ of each agent leaned by our models in TABLE \ref{tab:alpha_beta}. It could be observed that \textit{rank} of the confidences of each agent is relatively stable across different models. However, for each specific agent, the value of its confidence slightly increases when using UDA compared to the case without UDA. (e.g., the value of $\alpha$ of agent \textit{SR-SIM} increases from $0.700$ to $0.703$); Although the change of confidence is small in values, we have noted that this observation is consistent for all the agents, and as a result, the performance of the whole network improves with UDA. This indicates that the introduction of domain adaption could enable the whole network to better mine the knowledge of each agent.       

% Frankly, our methods take stock in each FR-IQA model with moderate confidence, because the distortion types in $\mathcal{X}_s$ is different from that of existing synthetically-distorted IQA dataset, $e.g.,$ LIVE \cite{sheikh2006statistical}, CSIQ \cite{larson2010most}, 
% where these FR-IQA methods perform excellent. After domain adaptation, small and stable gains of reliability are achieved in each model,  which corresponds to the performance improvement on target domain. This phenomenon motivates us we can boost model performance by employing more reliable IQA methods.

\subsubsection{gMAD competition}
In gMAD competition, our method first competes with the best $opinion$-$free$ knowledge-driven method - NIQE \cite{mittal2013making}, and representative gMAD pairs are shown in Fig. \ref{fig:koniq_niqe} and \ref{fig:spaq_niqe}, where target domain $\mathcal{X}_t$ are KonIQ-10k and SPAQ, respectively. From Fig. \ref{fig:koniq_niqe}(a) and (b), it could be observed that the images in the first row exhibit better perceptual quality than those in the second row (in agreement with our method while in disagreement with NIQE), indicating that our method correctly attacked NIQE by finding obvious counterexamples. When NIQE served as the attacker, our method successfully survived from the attack of NIQE, and the pairs of images found by the gMAD are of similar quality according to human perception. A similar observation was obtained when NIQE attacked our method when SPAQ was used in the target domain (see Fig. \ref{fig:spaq_niqe}). 

Our method was further compared to RankIQA \cite{liu2017rankiqa} using the gMAD competition. Fig. \ref{fig:koniq_rankiqa} shows the gMAD image pairs in the competition between RankIQA and our method adapted with KonIQ-10k as the target. Our method favored the first row images in Fig. \ref{fig:koniq_rankiqa}(a) and (b), which is consistent with human judgments, suggesting that our methods successfully attacked RankIQA. However, RankIQA failed to penalize the top image in Fig. \ref{fig:koniq_rankiqa}(c) and (d), which were spotted by our method.
Similarly, in Fig. \ref{fig:spaq_rankiqa} our method successfully identified strong failures of RankIQA  (see (a) and (b)), while RankIQA failed when switching their role (see Fig. \ref{fig:spaq_rankiqa}(c) and (d)). 

In a word, the proposed method, learning from synthetic data and multiple agents adapting to real-world distorted images, is proven to be an effective $opinion$-$free$ BIQA in the wild, obtaining the state-of-the-art performance with improved generalizability and robustness compared to other existing $opinion$-$free$ BIQA models.

\subsection {Ablation Study}
 A series of ablation experiments were conducted to test the effects of different components in our proposed method, $i.e.,$ distortion mixture, ADL, CLC, pixel-level domain mixup, and the number of agents.
\subsubsection {Effect of distortion mixture} Intuitively, how to simulate $\mathcal{X}_s$ could have an impact on final performance. To investigate this, we compare the performance of using the simulation of single distortion along with the mixture of several distortion types (see \ref{sec:synthesizing}). Technically, we build a control pseudo-labeled image set based on the reference image from the Waterloo Exploration Database \cite{ma2016waterloo}, where each image was only distorted by one type of distortions alone at five levels. Other settings, such as the generation of pseudo-labels, the rule to sample image pairs, the network structure, and hyperparameters of training, were unchanged. TABLE \ref{table:comparision} and TABLE \ref{table:ablation_single_mixture} report the quantitative comparison results with single and mixed distortions, and significant performance gains are obtained by the strategy of distortion mixture.

%\subsubsection{Effects of the number of annotators} 

\subsubsection{Effect of ADL} Binary cross-entropy (BCE) loss is one classical objective in UDA, where the easy-to-classify samples each contributes equally to each of the hard-to-classifier samples; however, in our study, we expect the hard-to-classifier examples to contribute more, $e.g.,$ to increase the weights of samples with authentic distortions that are far different from authentic distortions and reduce the weights when samples are easy to classified by domain discriminator. Thus, we introduce the ADL in Eq. (\ref{eq:flloss}) as the objective function of the domain classifier. TABLE \ref{table:ablation} depicts the performance comparison of BCE loss and the ADL in Eq. (\ref{eq:flloss}). As we can see, the ADL can yield subtle and stable performance improvement.   
\subsubsection {Effect of CLC} % Inspired by the approach offered in \cite{chen2020adversarial}, the ensemble of several different self-supervised tasks can offer additional performance gain. Diversity plays a key role in the final ensemble performance \cite{pang2019improving}.
The CLC was introduced in our model, which is different from the quality prediction task, as a complementary objective co-optimized with the quality prediction loss to help the rectification of feature extractor (see Fig. \ref{fig:pipeline}). We test the performance of our model, with and without CLC; results are presented in TABLE \ref{table:ablation}. It can be observed that there is an obvious performance gain after the introduction of the CLC, indicating the CLC is effective in our model and could boost the performance.

\subsubsection {Effect of domain mixup} % Generally, mixup \cite{xu2020adversarial} is a data augmentation technique, generating convex combinations of additional training samples with their corresponding labels. Motivated by this, 
We further test the effect of domain pixel-level mixup regularization. TABLE \ref{table:ablation} shows the difference between with/without mixup-based regularization. We observe a small performance gain achieved by incorporating a mixup strategy. This gain might attribute to the fact that the mixup operation could augment more mixed instances within each domain, thus helping the model to learn meaningful internal feature representations in the latent space.

\renewcommand{\multirowsetup}{\centering}
\begin{table}[t]
    \caption{Effect of the number of agents. $\mathcal{X}_t$ is included in the bracket. The top two correlations obtained by ours are highlighted in boldface}
    \label{table:effect_of_num}
    \vspace{-.3cm}
	\begin{center}
		\begin{tabular}{l|cccc}
    		\toprule[1pt]
			\multirow{2}{*}{SRCC} & \# of & LIVE & KonIQ-10k & SPAQ \\
			& agents & \cite{sheikh2006statistical} & \cite{hosu2020koniq} & \cite{fang2020cvpr} \\
			\hline
			%$-$ & $0.914$ & $0.618$ & $0.715$\\
			\multirow{4}{*}{Ours(KonIQ-10k)} & $1$ & $0.887$ & $0.657$ & $0.821$\\
			& $2$ & $0.890$ & $0.665$ & $0.820$\\
		    & $4$ & $0.908$ & $0.682$ & $0.822$\\
			& $6$ & $0.907$ & $\textbf{0.717}$ & $0.826$\\
			\hline
			\multirow{4}{*}{Ours(SPAQ)} & $1$ & $0.876$ & $0.654$ & $0.825$ \\
			& $2$ & $0.901$ & $0.660$ & $0.825$\\
			& $4$ & $\textbf{0.913}$ & $0.675$ & $\textbf{0.828}$\\
			& $6$ & $\textbf{0.920}$ & $\textbf{0.712}$ & $\textbf{0.838}$\\
			\hline
			\hline
			\multirow{2}{*}{PLCC} & \# of & \multirow{2}{*}{LIVE} & \multirow{2}{*}{KonIQ-10k} & \multirow{2}{*}{SPAQ}\\
			& agents & \\
			\hline
			\multirow{4}{*}{Ours(KonIQ-10k)} & $1$ & $0.881$ & $0.679$ & $0.825$\\
			& $2$ & $0.890$ & $0.682$ & $0.826$\\
		    & $4$ & $0.904$ & $0.711$ & $0.825$\\
			& $6$ & $\textbf{0.906}$ & $\textbf{0.740}$ & $0.831$\\
			\hline
			\multirow{4}{*}{Ours(SPAQ)} & $1$ & $0.870$ & $0.680$ & $0.828$\\
			& $2$ & $0.898$ & $0.676$ & $0.826$\\
			& $4$ & $0.906$ & $0.698$ & $\textbf{0.837}$\\
			& $6$ & $\textbf{0.912}$ & $\textbf{0.736}$ & $\textbf{0.844}$\\
			\bottomrule[1pt]
		\end{tabular}
	\end{center}
\end{table}

\subsubsection {Effect of the number of agents} Last, we study the influence of the number of agents ($i.e.$, FR-IQA models). We vary the number of agents systematically with values set to  $1$, $2$, $4$, $6$ and test the performance of our method. For each of the agent numbers at $1$, $2$, $4$, we repeated the experiments by randomly selecting three different combinations of agents and report the mean values in TABLE \ref{table:effect_of_num}. It could be observed that, in most of the cases, with the increase of the number of FR-IQA agents used, the performances on KonIQ-10k and SPAQ are improved, and the model with six agents achieved the relatively best performance.
Therefore, it is possible that the model performance could be further improved by employing more diverse and reliable FR-IQA models.

\section{Conclusion}
Collecting large-scale human-rated IQA  image datasets is prohibitively labor-expensive and costly. Existing $opinion$-$free$ BIQA models were mainly trained and tested on synthetically-distorted images and generalize poorly to authentically-distorted images in the wild. To tackle this issue, in this work, we present a simple and yet effective $opinion$-$free$ BIQA model to assess the quality of images captured in the wild. Our proposed method demonstrated superior performance compared to existing ones. It has a lot of practical significance, especially for building task-specific BIQA \cite{zhang2019ranksrgan}, where general BIQA models do not perform well and human labels are hard to collect in those tasks, $e.g.,$ perceptual optimization for colorization, super-resolution, restoration, enhancement, $etc$. Our proposed method can be potentially improved by (1) incorporating more advanced FR-IQA models and UDA methods; (2) adopting ideal distortion simulation algorithms to synthesize authentic distortions. 

% , which we believe arises from (1) the simulation of authentic distortions; (2) a suitable and efficient UDA method to reduce the domain gap between the synthetically-distorted images and authentically-distorted images captured in the wild. 
%/******************/
% We conduct our experiment on two relative large-scale human-rated real-world IQA datasets, and the   quantitative results in term of SRCC and PLCC on these two real-world datasets prove state-of-the-art performance compared with existing $opinion$-$free$ BIQA methods, and the qualitative results from the gMAD competition verify the generalizability and robustness of our proposed method.
%Although, the performance of our proposed methods is still inferior to models trained on fully human-rated IQA datasets \textbf{(JZ:do we have experiments to demo this? if not, then delete.)}.  \textbf{(if possible, then delete.)}. 
%/******************/

% use mization for restoration, hsection* for acknowledgment
\section*{Acknowledgment}
The authors would like to thank Hanwei Zhu from City University of Hong Kong, Xuelin Liu from Jiangxi University of Finance and Economics%, Mingming Gong from The University of Melbourne 
for fruitful discussions throughout the development of this work.

\ifCLASSOPTIONcaptionsoff
  \newpage
\fi



%\bibliographystyle{IEEEtran}
% argument is your BibTeX string definitions and bibliography dataset(s)
{\small
\bibliographystyle{ieee_fullname}
\bibliography{main}
}

\end{document}